\documentclass[conference]{IEEEtran}
\usepackage{fancyhdr}
\IEEEoverridecommandlockouts
\usepackage{cite}
\usepackage{fontawesome}
\usepackage{amsmath,amsfonts,amscd}

\usepackage{amssymb}
\usepackage{fancyhdr}
\usepackage{mathrsfs}
\usepackage{graphicx}
\usepackage{textcomp}
\usepackage{multirow}
\usepackage{amsthm}
\usepackage[T1]{fontenc}
\usepackage{enumitem}
\usepackage{algorithm,algorithmicx,algpseudocode}
\usepackage{listings}
\usepackage{pifont}
\usepackage{booktabs}
\usepackage{threeparttable}
\usepackage{color}
\usepackage{fontawesome}

\AtBeginDocument{%
  \providecommand\BibTeX{{%
    \normalfont B\kern-0.5em{\scshape i\kern-0.25em b}\kern-0.8em\TeX}}}

\newcommand{\one}{\ding{182}}
\newcommand{\two}{\ding{183}}
\newcommand{\three}{\ding{184}}
\newcommand{\four}{\ding{185}}
\newcommand{\five}{\ding{186}}
\newcommand{\six}{\ding{187}}
\newcommand{\seven}{\ding{188}}

\definecolor{r1}{RGB}{87,114,158}
\definecolor{r2}{RGB}{204,137,99}
\definecolor{r3}{RGB}{93,157,107}
\definecolor{r4}{RGB}{129,114,180}
\definecolor{r5}{RGB}{147,120,96}
\usepackage{marginnote}
\usepackage{soul} 
\usepackage{hhline}
\usepackage{ulem}
\definecolor{lavenderblue}{rgb}{0.8, 0.8, 1.0}
\sethlcolor{lavenderblue}

\usepackage{xparse}

\usepackage{mdframed}
\global\mdfdefinestyle{review}{%
linecolor=lavenderblue,linewidth=3pt,%
leftmargin=0cm,rightmargin=0cm,%
skipabove=0cm,skipbelow=0cm,%
innerrightmargin=0cm,innerleftmargin=0cm,%
innerbottommargin=0cm,innertopmargin=0cm,%
backgroundcolor=lavenderblue
}
\global\mdfdefinestyle{reviewtext}{%
linecolor=lavenderblue,linewidth=0pt,%
leftmargin=0cm,rightmargin=0cm,%
skipabove=0.1cm,skipbelow=0.1cm,%
innerrightmargin=0cm,innerleftmargin=0cm,%
innerbottommargin=0cm,innertopmargin=0cm,%
backgroundcolor=lavenderblue
}

\def\BibTeX{{\rm B\kern-.05em{\sc i\kern-.025em b}\kern-.08em
    T\kern-.1667em\lower.7ex\hbox{E}\kern-.125emX}}
\begin{document}

\DeclareDocumentCommand\review{m g g}{%
{\IfNoValueF {#2}{%
\IfNoValueF {#3}{%
{\marginnote{\sethlcolor{#3}\hl{\normalfont \textbf{{\normalsize{\color{white}#2}}}}}%
}%
}%
\IfNoValueT {#3}{%
{\marginnote{\normalfont \textbf{\normalsize{#2}}}%
}%
}%
}%
\hl{#1}%
}%
}

\title{Efficient Quantized Sparse Matrix Operations on Tensor Cores}

\author{\IEEEauthorblockN{Shigang Li}
\IEEEauthorblockA{School of Computer Science \\
Beijing University of Posts and Telecommunications; \\
Department of Computer Science \\
ETH Zurich \\
shigangli.cs@gmail.com}
\and
\IEEEauthorblockN{Kazuki Osawa}
\IEEEauthorblockA{Department of Computer Science \\
ETH Zurich \\
kazuki.osawa@inf.ethz.ch}
\and
\IEEEauthorblockN{Torsten Hoefler}
\IEEEauthorblockA{Department of Computer Science \\
ETH Zurich \\ 
htor@inf.ethz.ch}
}

\maketitle
\thispagestyle{fancy}
\lhead{}
\rhead{}
\chead{}
\lfoot{\footnotesize{SC22, November 13-18, 2022, Dallas, Texas, USA}}
\rfoot{}
\cfoot{}
\renewcommand{\headrulewidth}{0pt}
\renewcommand{\footrulewidth}{0pt}

\begin{abstract}
The exponentially growing model size drives the continued success of deep learning, but it brings prohibitive computation and memory cost. 
From the algorithm perspective, model sparsification and quantization have been studied to alleviate the problem. From the architecture perspective, hardware vendors provide Tensor cores for acceleration.
However, it is very challenging to gain practical speedups from sparse, low-precision matrix operations on Tensor cores, because of the strict requirements for data layout and lack of support for efficiently manipulating the low-precision integers. 
We propose \textit{Magicube}, a high-performance sparse-matrix library for low-precision integers on Tensor cores. 
Magicube supports SpMM and SDDMM, two major sparse operations in deep learning with mixed precision.
Experimental results on an NVIDIA A100 GPU show that Magicube achieves on average 1.44x (up to 2.37x) speedup over the vendor-optimized library for sparse kernels, and 1.43x speedup over the state-of-the-art with a comparable accuracy for end-to-end sparse Transformer inference.

\end{abstract}


\section{Introduction}
\label{sec:intro}

Recent progress in state-of-the-art deep learning has been driven by the increasing scale of computation, data, and models, and this scaling trend is expected to continue~\cite{hestness2017deep,rosenfeld2019constructive,kaplan2020scaling,brown2020language}.
Such large-scale deep learning models require large amounts of energy and carbon emissions for training~\cite{openai2018aiandcompute,strubell2019energy,patterson2021carbon}, and evaluating inferences with limited computational and memory resources is challenging.
The main techniques to reduce the memory footprint and inference latency are sparsification and quantization of matrix operations~\cite{gholami2021survey,hoefler2021sparsity}.  

While these compression methods theoretically reduce the number of operations, speedup each operation, and improve memory bandwidth, it is not easy to obtain practical speedups on accelerators.
In deep learning, the sparsity of the matrix that can be achieved while preserving the prediction accuracy of the model is relatively small (e.g., 50-90\%)~\cite{hoefler2021sparsity}. 
Therefore, with sparse kernels that target high sparsity (e.g., >\,99\%) provided by e.g., cuSPARSE~\cite{cusparse}, it is difficult to exceed the performance of the dense counterparts (e.g., cuBLAS).
To obtain practical speedups with accelerators, cuSPARSELt~\cite{cusparselt} utilizes Tensor Cores sparsity \cite{SparseTC} and achieves the double peak performance compared to the dense counterparts in several low-precision datatypes (e.g., fp16, int8, int4). Yet, this library imposes strict constraints on the data layout (i.e., 2:4 structured sparsity) with sparsity constrained to 50\%.
Gale et al.~\cite{gale2020sparse} proposed Sputnik, a library for sparse matrix-matrix multiplication (SpMM) and sampled dense-dense matrix multiplication (SDDMM) in fp32 and fp16 datatypes that takes advantage of the properties of matrices in deep learning (e.g., a high number of nonzeros per row) and works with a fine-grained sparse data layout. 
Sputnik outperforms cuSPARSE on fp32 deep learning workloads at moderate sparsity (e.g., 70\%) on NVIDIA V100 GPUs.
Chen et al.~\cite{chen2021efficient} pointed out that the existing sparse kernels cannot realize speedups over the dense counterparts when low-precision is used due to the lack of data reuse. 
Also, they showed that the SpMM kernel for block sparse matrix multiplication in cuSPARSE requres the block size to be larger than 8 to achieve speedup. This makes it more challenging to keep the model accuracy. 
To address these issues, they proposed vectorSparse, a library using a sparse encoding with dense 1-D block of shape e.g., $8\times 1$, $4\times 1$, that improves the data reuse in fp16 while maintaining the flexibility in data layout.

\newcommand{\cmark}{{\color{green}\ding{51}}}
\newcommand{\xmark}{{\color{red}\ding{55}}}
\newcommand{\good}{\faThumbsOUp}
\newcommand{\bad}{\faThumbsDown}

\setlength{\tabcolsep}{2pt}
\begin{table}[]
    \centering
    \caption{Supported input operand (low-)precision and sparsity constraints in various sparse-matrix libraries. TC: whether it runs on Tensor cores.}
    \begin{tabular}{cccccccc}
    \toprule
    \multirow{2}[2]{*}{Library} & \multicolumn{4}{c}{Precision} & \multicolumn{2}{c}{Sparsity} & \multirow{2}[2]{*}{TC} \\
    \cmidrule(lr){2-5} \cmidrule(lr){6-7}
    & fp16 & int8 & int4 & mixed & granularity & DL-friendly? \\
    \midrule
    \multirow{2}{*}{cuSPARSE~\cite{cusparse}} & \cmark & \cmark & \xmark & \xmark & fine-grained & \bad & \bad \\
    & \cmark & \cmark & \xmark & \xmark & block & \good & \good \\
    cuSPARSELt~\cite{cusparselt} & \cmark & \cmark & \cmark & \xmark & 2:4 structured & \good & \good \\
    Sputnik \cite{gale2020sparse} & \cmark & \xmark & \xmark & \xmark & fine-grained & \good & \bad \\
    vectorSparse \cite{chen2021efficient} & \cmark & \xmark & \xmark & \xmark & 1-D block & \good & \good \\
    {\bf Magicube} & \xmark & \cmark & \cmark & \cmark & 1-D block & \good & \good \\
    \bottomrule
    \end{tabular}
    \label{tab:sparse-mat-lib}
\end{table}

These point results are part of a bigger question: which combinations of low-precision datatypes and sparsity should be supported in hardware and which others can be supported in software. 
Combining sparsity and quantization has been shown to be highly effective in deep learning~\cite{van2020bayesian}. 
Yet, both have different tradeoffs, and hardware vendors must choose a configuration in this costly design space for each product. 
In addition to this, due to the lack of support for efficiently manipulating low-precision integers (e.g., int4), it is challenging to realize high-performance sparse and quantized matrix operations on Tensor cores~\cite{nvidia2017nvidia,jia2018dissecting,amperetuning}.
In our work, we define two additional software design points for NVIDIA A100 tensor cores with our \emph{Magicube}\footnote{the name is inspired by the similarity of quantization to the Rubik's cube, and the similarity of tensor core data marshalling to playing the Rubik's cube.} library: (1) small 1-D block sparse operations with low precision integer types, (2) small 1-D block sparse operations with mixed precision. Note that, in this work, we define the \textit{mixed precision} as the two input matrices of matrix multiplication have different precision.
In deep learning, assigning data types of different precision to different types of quantities (e.g., weight, activation) with different sensitivities to quantization has proven to be effective both in terms of reducing accuracy degradation and in terms of hardware efficiency~\cite{gholami2021survey}.
Table~\ref{tab:sparse-mat-lib} summarizes the state-of-the-art sparse matrix libraries for various datatypes on GPUs.

We consistently outperform all existing libraries by more than 40\% for those type combinations by efficiently marshalling the inputs to tensor cores and emulating low and mixed precision integer operations algebraically. We also show that those optimizations translate to end-to-end performance improvements of more than 40\% for transformer networks, the most promising candidate for today's and future large-scale deep learning systems~\cite{brown2020language}. Our main contributions are:
\begin{itemize}
 \item We design a sparse matrix format SR-BCRS which is friendly to low-precision integers on Tensor cores.
 \item We introduce highly-optimized SpMM and SDDMM kernels. Specifically, we propose an novel online transpose strategy to efficiently manipulate fine-grained data and meet the data layout requirements. 
 \item Magicube supports mixed precision using efficient algebraic type emulation, in which the utilization of Tensor cores is improved through operations stacking.
 \item Magicube achieves significant speedup over the latest dense and sparse library for both microbenchmarks and a real-world deep learning application. The model accuracy is also verified.
\end{itemize}

We evaluate the performance of sparse kernels over 1,536 sparse matrices with different sizes and sparsity. Results on NVIDIA A100 GPU shows that Magicube achieves on average (geometric mean) 1.44x speedup for SpMM over cuSPARSE. For end-to-end sparse Transformer \cite{child2019generating} inference, Magicube achieves 1.43x speedup over vectorSparse~\cite{chen2021efficient} (the state-of-the-art sparse library with fp16 on Tensor cores) and 1.50x speedup over PyTorch with cuDNN (the fp16 dense library), with a comparable accuracy. The source code of Magicube is available at \textcolor{blue}{https://github.com/Shigangli/Magicube}

\section{Background and Related Work}
\label{sec:background}

\subsection{Compression in Deep Learning}
Sparsification and quantization are the common ways to compress deep neural networks for reducing energy and performance costs for training and inference \cite{tung2018deep, yang2020automatic, yu2020joint, wang2020apq, srivastava2019joint}.
Sparsification reduces the number of operations in workloads ({e.g.}, matrix-matrix multiplication, convolution) by ignoring redundant elements in the operands that contribute little to learning and prediction \cite{han2015deep, you2019drawing, hoefler2021sparsity}.
Quantization, on the other hand, speeds up each operation and improves the memory bandwidth by representing the operands with low bits, such as fp16, 8-bit, and 4-bit integers \cite{rastegari2016xnor, micikevicius2017mixed, wu2020integer, hubara2021accurate, nagel2021white, gholami2021survey}.
Both serve to reduce the storage requirement by compressing neural network weights, inputs, and intermediate representations (activation and backpropagated error).

In recent years, Transformer models \cite{vaswani2017attention} based on the attention mechanism \cite{bahdanau2014neural} are becoming mainstream in various domain applications, such as natural language processing and computer vision.
Since the effectiveness of the pre-training and fine tuning paradigm using large Transformer models ({e.g.}, BERT\cite{devlin2018bert}, GPT-3\cite{brown2020language}) has been demonstrated, the importance of reducing amounts of carbon emission, energy consumption and computational and memory costs in training and inference has been increasing \cite{openai2018aiandcompute,strubell2019energy,patterson2021carbon}.
To reduce the memory footprint and inference latency, weight pruning \cite{sanh2020movement,lagunas2021block,mao2021tprune} and quantization \cite{bhandare2019efficient, zafrir2019q8bert, mao2020ladabert, kim2021bert} for giant Transformer models have been studied. 
Sparsification of the attention map \cite{child2019generating,wang2020linformer,zaheer2020big,beltagy2020longformer,choromanski2020rethinking} has also been studied to reduce the computational and memory complexity of the self-attention which is proportional to the square of the sequence length.

\subsection{Optimization for Sparse and Quantized Operations}
Compression induces sparse and quantized operations, but appropriate hardware and/or a series of performance optimization is required to gain practical speedups~\cite{han2016eie, gale2020sparse, chen2021efficient, li2019bstc, li2020accelerating, feng2021apnn}.
Performance optimization for sparse matrix operations in scientific computing has been studied~\cite{wang2018swsptrsv, chen2018performance, xie2021fast, xie2019ia, niu2021tilespmv, xie2021pattern}.
However, the sparsity of matrices assumed in these domains usually exceeds 99\%, whereas in deep learning, it is typically around 50-90\% to maintain the prediction accuracy of the neural network~\cite{hoefler2021sparsity}.
Therefore, it is more challenging to achieve practical speedups over the dense counterpart for deep learning workloads.

AI accelerators, such as Tensor cores~\cite{nvidia2017nvidia,jia2018dissecting,amperetuning}, bring unprecedented performance for deep learning workloads with low precision (floating- and fixed-point). But this is mainly for dense matrix multiplications. Structured sparsity is also natively supported on Tensor cores~\cite{SparseTC}, but it has strict requirements for the distribution of non-zero elements, e.g., 50\% sparsity ratio, and sparsity patterns, e.g., 1:2 or 2:4~\cite{nvptx, SparseTC}, which may limit its generality and usability. Gale et al.~\cite{gale2020sparse} introduce Sputnik
that optimizes the performance of deep learning workloads with more general fine-grained sparsity on CUDA cores, and outperforms cuSPARSE with relatively low sparsity. Chen et al.~\cite{chen2021efficient} propose vectorSparse to improve the performance of structured sparsity (with less constrains) on Tensor cores. But both Sputnik and vectorSparse target on sparse workloads in half precision. Different from previous work, we focus on quantized sparse matrix operations (SpMM and SDDMM) with low-prevision integers, and present excellent performance.

\section{Tensor cores of NVIDIA GPU and data layout for low-precision integers}

NVIDIA GPUs consist of an array of streaming multiprocessors (SMs) and each SM contains small processing components (such as CUDA cores and Tensor cores). The CUDA (NVIDIA GPU programming model) kernel is executed with multithreading. Threads in GPU kernels are organized by a grid of thread blocks, and each thread block consists of warps (each warp has 32 threads). Warps are the basic scheduling units in CUDA. 

Tensor Core Unit (TCU)~\cite{nvidia2017nvidia} has been added to the NVIDIA GPUs since the Volta architecture~\cite{nvidia2017nvidia}, which is designed specifically for deep learning and provides 8× peak FLOPs (pf16) than FPU on CUDA cores. Tensor cores on newer architectures, such as Ampere~\cite{amperetuning} and Hopper~\cite{nvhopper}, add support for matrix-matrix multiplication with low-precision integers (8-bit, 4-bit, and even 1-bit), which provides double, quadruple, or even higher peak performance than fp16. As shown in Table~\ref{tab:tensor-cores}, Tensor cores provide almost all the computational power of low precision integers on NVIDIA GPUs. Therefore, we target on Tensor cores to optimize the quantized sparse kernels for deep learning. In the following we use \texttt{int}$x$ to represent integers with $x$ bits. To program on Tensor cores, CUDA provides warp-level APIs with the semantic of Matrix Multiply-Accumulate (MMA), in which a warp of 32 threads collaboratively execute one or several dense matrix multiplications and accumulate the outputs. From the programming perspective, CUDA provides two APIs with the semantic of MMA, including \texttt{WMMA} in (high-level) C++ and \texttt{mma} in (low-level) NVPTX (quasi-assembly language). In Magicube, we use the \texttt{mma} APIs. Dense operations with extremely low precision (int2 or even int1) for deep learning have been studied in~\cite{li2020accelerating, li2019bstc, feng2021apnn}. In this work, we focus on moderate precision (e.g., int4, int8) with sparsity.

\setlength{\tabcolsep}{3pt}
\begin{table}[]
    \centering
    \caption{The total peak TFLOPS/TOPS (Tensor cores $+$ CUDA cores) and the percentage of Tensor cores in total }
    \begin{tabular}{cccccc}
    \toprule
    GPU & fp16 & int8 & int4 \\
    \midrule
    V100 & 126\,TFLOPS (88.9\%) & - & - \\
    A100 & 390\,TFLOPS (80\%) & 702\,TOPS (88.9\%) & 1,248\,TOPS (100\%) \\
    H100 & 1,120\,TFLOPS (89.2\%) & 1,696\,TOPS (94.3\%) & - \\
    \bottomrule
    \end{tabular}
    \label{tab:tensor-cores}
\end{table}

\begin{figure}[ht!]
\centering\includegraphics[width=0.99\linewidth]{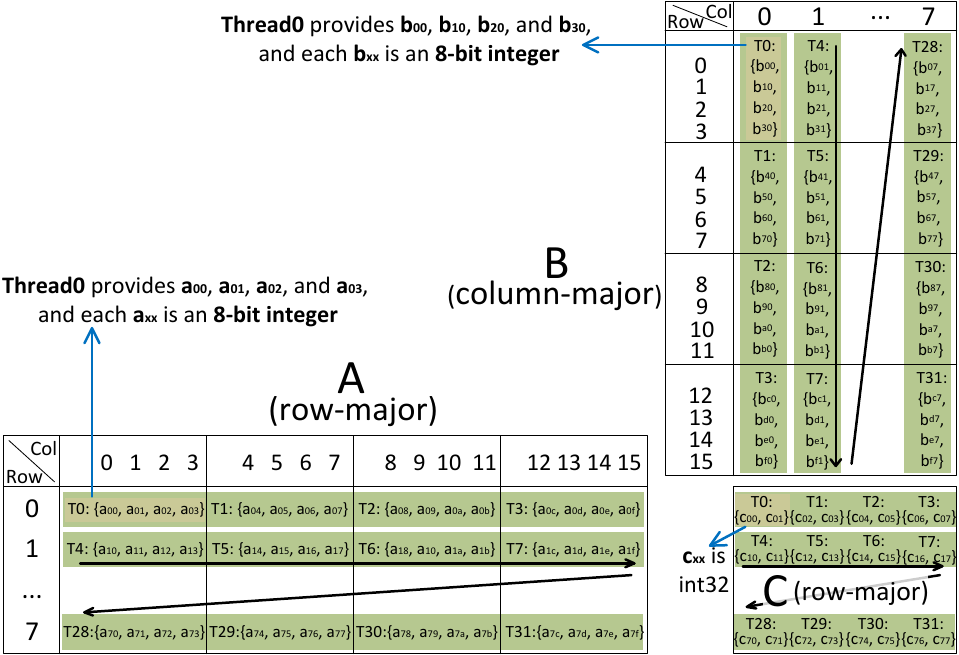}
\caption{\label{fig:m8n8k16} Data layout of int8 \texttt{mma} with the shape of \texttt{m8n8k16}.}
\end{figure}

\setlength{\tabcolsep}{3pt}
\begin{table}[]
    \centering
    \caption{Matrix shapes for \texttt{mma} on Tensor cores}
    \begin{tabular}{p{2cm}p{3.9cm}}
    \toprule
    Precision & Supported shapes \\
    \midrule
    int4/uint4 &  \textbf{\textit{m8n8k32}},\ \  \textit{m16n8k32}, \ \ \textit{m16n8k64}\\
    int8/uint8 &  \textbf{\textit{m8n8k16}},\ \  \textit{m16n8k16}, \ \ \textit{m16n8k32}\\

    \bottomrule
    \end{tabular}
    \label{tab:shapes}
\end{table}

On Tensor cores, several shapes of \texttt{mma} are supported for each precision. The supported shapes for int4 and int8 are shown in Table~\ref{tab:shapes}. See~\cite{nvptx} for the supported shapes of other precision. In Magicube we choose to use the smallest shapes (highlighted in Table~\ref{tab:shapes}) to exploit the performance with small sparsity granularity, since smaller granularity usually achieves better accuracy results under the same sparsity ratio~\cite{mao2017exploring,chen2021efficient}. For \texttt{mma} with int8 and int4, we use the shapes of \textit{m8n8k16} and \textit{m8n8k32}, respectively. The data layout of \texttt{mma} with int8 in shape \textit{m8n8k16} is shown in Figure~\ref{fig:m8n8k16}. The shape of the output matrix C (each element is an int32) is \textit{m}*\textit{n} (i.e., 8*8), and the reduction dimension is \textit{k} = 16. The shape of the Left-Hand-Side (LHS) matrix A is 8*16 and the shape of the Right-Hand-Side (RHS) matrix B is 16*8, and each element of A and B is an int8. As shown in Figure~\ref{fig:m8n8k16}, the elements of A, B and C are uniformly distributed among the registers of a warp (32 threads). Note that A must be row-major and B must be column-major. Each thread provides 4 int8 elements to both A and B. Programmers have to decompose the whole workload into small MMAs (e.g., \textit{m8n8k16}), and match the restrict requirement for the data layout of A, B and C. The data layout of \texttt{mma} with int4 in shape \textit{m8n8k32} is similar to int8 in shape \textit{m8n8k16}, except that each element of A and B is an int4, the reduction dimension k increases to 32, and each thread provides 8 int4 elements to both A and B.

\section{Implementation and optimization for sparse matrix operations in deep learning}

\subsection{Sparse matrix format}

\begin{figure}[ht!]
\centering\includegraphics[width=0.9\linewidth]{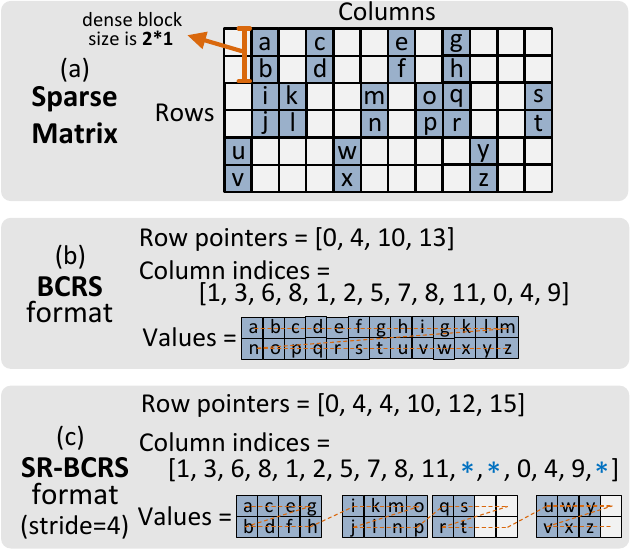}
\caption{\label{fig:sparseformat} BCRS format vs SR-BCRS format.}
\end{figure}

Sparse matrices are important workloads in both the scientific computing~\cite{vuduc2005oski, wang2018swsptrsv, chen2018performance, xie2021fast, xie2019ia, liu2021sparta} and deep learning~\cite{gale2020sparse, chen2021efficient, child2019generating, zaheer2020big, hoefler2021sparsity} domains. The Compressed Row Storage (CRS) format is the most popular method to compress sparse matrices because of its simplicity. The Block Compressed Row Storage (BCRS)~\cite{pinar1999improving} format is further proposed to improve the data reuse on L1 cache or registers. The column vector sparse encoding used in vectorSparse~\cite{chen2021efficient} is a special case of BCRS, in which each dense block is an 1-D block. An example of BCRS with 1-D blocks is shown in Figure~\ref{fig:sparseformat}.

Since BCRS with 1-D dense blocks is sufficient to exploit data reuse in the sparse deep learning workloads~\cite{chen2021efficient} and small sparsity granularity is good for model accuracy~\cite{gale2020sparse}, we also focus on structured sparsity with 1-D dense blocks rather than 2-D dense blocks. Different from previous work, to exploit the performance of the sparse workloads with low-precision integers on Tensor cores, we propose to use a \textit{S}trided \textit{R}ow-major BCRS (SR-BCRS) format. Now we present the details of SR-BCRS by comparing it with BCRS format. BCRS consists of row pointers, column indices for the dense (non-zero) vectors, and the dense vectors stored in a consecutive array. In contrast, the dense vectors in SR-BCRS are stored in a stride-wise row-major manner, as shown at the bottom of Figure~\ref{fig:sparseformat}. Zero values are padded for the last stride of a vector row if the number of dense vectors in the row is not a multiple-of-stride. Correspondingly, the column indices are also padded with invalid values (*). Here the size of the stride is equal to the reduction (i.e., $k$) dimension of the \texttt{mma} operation we use. For example, \texttt{stride} = 16 for \texttt{mma} with int8. In this way, the threads in a warp can consecutively load the data of the LHS matrix to the registers and the data layout requirement (as shown in Figure~\ref{fig:m8n8k16}) is directly satisfied. Here the length of the 1-D dense block ($V$) supported in SR-BCRS is <= 8 (i.e., the $m$ dimension of \texttt{mma}). If $V$ = 8, the \texttt{mma} operation is fully utilized; if $V$ = 4, the utilization of \texttt{mma} is 50\%. To support this strided format, we need 2$M$ ($M$ is the number of rows of a sparse matrix) row pointers. For each row, we need one pointer to indicate the address of the first dense vector and another to indicate the last dense vector in the current row of vectors.

\begin{figure}[ht!]
\centering\includegraphics[width=0.99\linewidth]{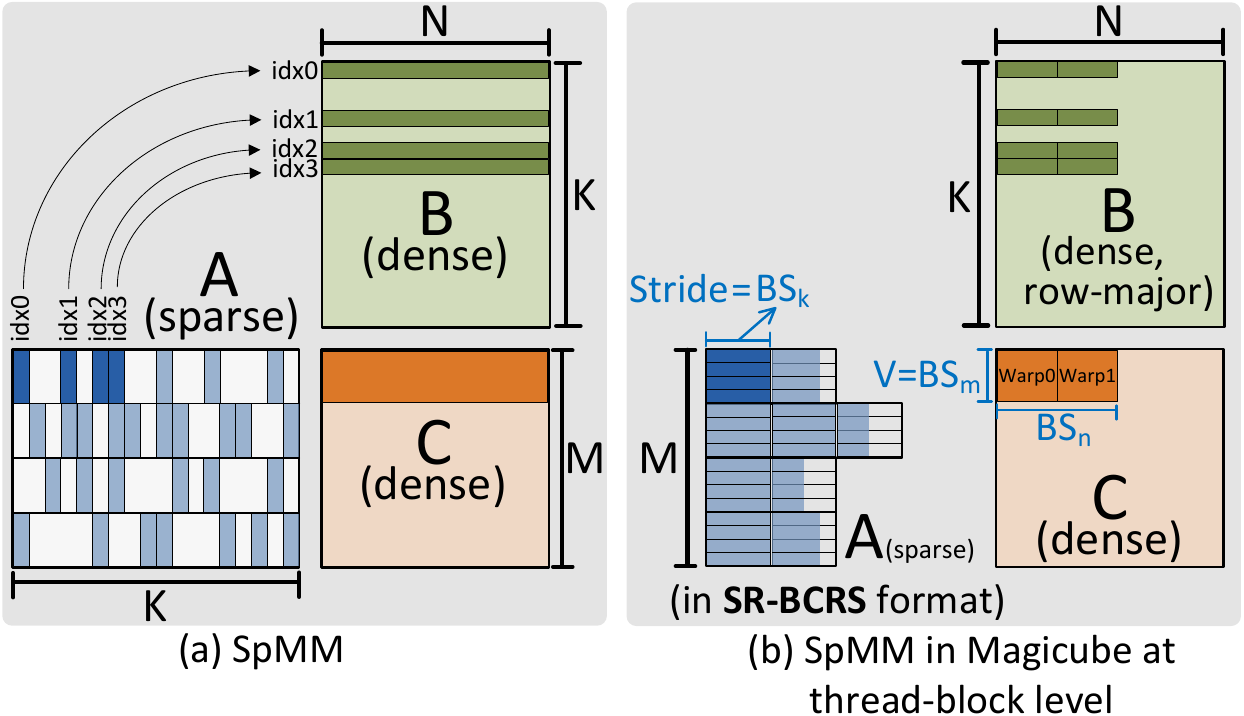}
\caption{\label{fig:spmmblocks} SpMM and its thread-block view in Magicube.}
\end{figure}
\subsection{SpMM in Magicube}
\label{sec:spmm}

Sparse matrix-matrix multiplication (SpMM) is a major sparse workload in deep learning. For example, in the forward pass of a pruned model, the sparse weight matrix will be multiplied by a dense activation matrix. In sparse transformers \cite{child2019generating}, the self-attention is calculated by multiplying a sparse attention weight matrix by a dense value matrix. These all result in an SpMM operation. Figure~\ref{fig:spmmblocks}(a) shows an example of SpMM with structured sparsity (i.e., 1-D blocks), in which matrix A can be recognized as a pruned weight matrix or sparse attention mask with structured sparsity.
Note that the column indices of the dense vectors are used to load the corresponding rows of the dense matrix B.

\subsubsection{Thread blocks of SpMM}

Figure~\ref{fig:spmmblocks}(b) shows the implementation of SpMM in Magicube at the thread-block level. Since we focus on quantized sparse matrix operations, here we assume each element in A and B is int8. Suppose we have 2 warps in a thread block. Each thread block is responsible for an output block of size $BS_m$*$BS_n$. Here $BS_m$=$V$ (i.e., vector length). $BS_m$ can be set to a multiple-of-$V$, but this does not help to improve the data reuse since each row of vectors may have different column indices. In each step, each thread block calculates the partial results with the reduction dimension $BS_k$. Here $BS_k$ is equal to the stride size in the SR-BCRS format, and also equal to the reduction dimension of \texttt{mma}. Overall, each thread block needs \textit{nnz}/$BS_k$ steps (partial results are accumulated) to obtain the final results. Benefiting from the SR-BCRS format, the data layout requirement of the LHS matrix of \texttt{mma} is directly satisfied by consecutively loading data in a stride. The LHS matrix is loaded in a coalesced way from global memory to shared memory, and the data are shared among warps.

\begin{figure}[ht!]
\centering\includegraphics[width=0.99\linewidth]{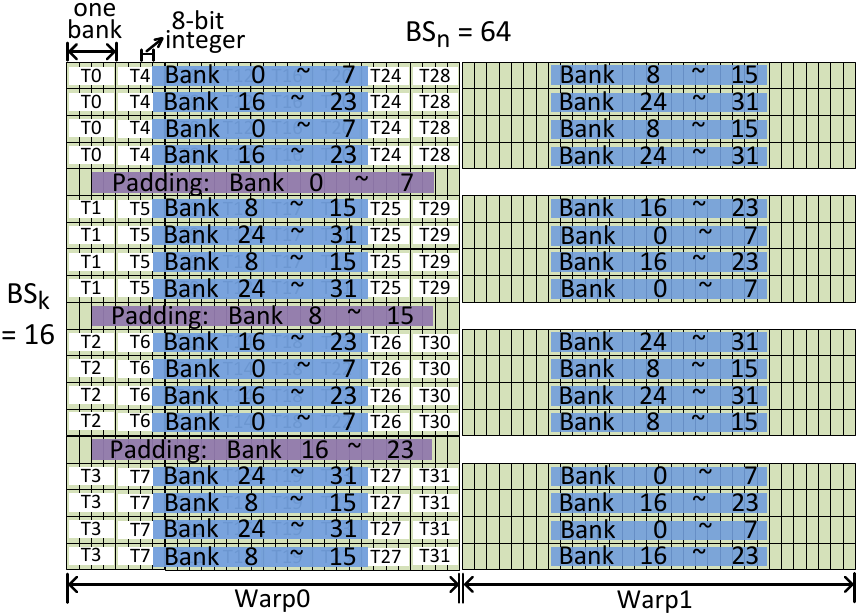}
\caption{\label{fig:spmmsm} Conflict-free accesses for the RHS matrix of SpMM with int8 on shared memory.}
\end{figure}

\begin{figure}[ht!]
\centering\includegraphics[width=0.99\linewidth]{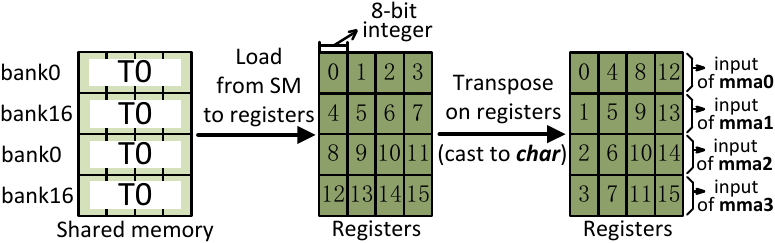}
\caption{\label{fig:spmmreg} Local transpose on registers for RHS matrix of SpMM with int8.}
\end{figure}

\subsubsection{Efficient online transpose for dense matrix B with int8}\label{sec:transpose}
\begin{figure}[ht!]
\centering\includegraphics[width=0.99\linewidth]{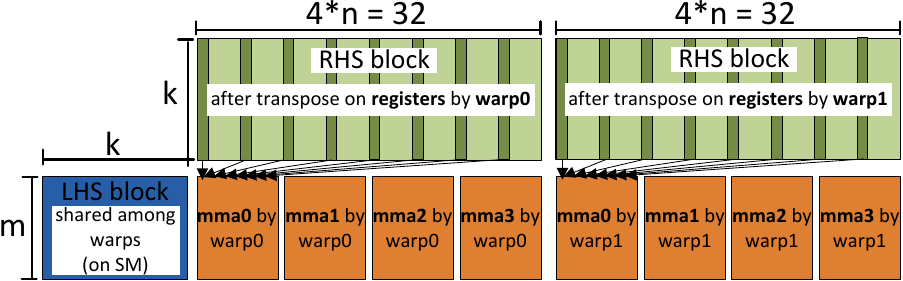}
\caption{\label{fig:spmmwarps} The warp-level view of the MMAs in SpMM with int8.}
\end{figure}

Next, we discuss how to feed the data block of B to the RHS matrix of \texttt{mma}. Here we use row-major storage for the dense matrix B. However, recall that the RHS matrix of \texttt{mma} must be in column-major, there is a mismatch for the layout. Transposing B to column-major beforehand does NOT help since the column indices of the dense vectors are not consecutive. Therefore, we propose an efficient online transpose strategy for the row-major blocks of B, which contains three major steps. First, the threads in a block collaboratively load rows of B from global memory to padded buffers on shared memory. An example with $BS_n = 64$ is shown in Figure~\ref{fig:spmmsm}. Each row of 64 int8 values is coalesced into a single 64B memory transaction~\cite{cudaguide}, boosting the efficiency of global memory access. Note that we may set $BS_n = 128$ to coalesce to an 128B memory transaction~\cite{cudaguide} for higher efficiency for large $N$. Second, each thread will in turn load 4 int32 items from shared memory to local registers. The items accessed by each thread of warp0 are labelled in Figure~\ref{fig:spmmsm}. On NVIDIA GPUs, shared memory are partitioned into banks and each bank has a width of 32 bits. Threads in a warp accessing different banks can be served by one cycle; otherwise threads in a warp accessing different banks incurs bank conflicts which degrade the performance. By padding 8 int32 items after every 64 int32 items, there is no bank conflict within each warp. Third, each thread conducts a transpose on local registers with a granularity of int8, as shown in Figure~\ref{fig:spmmreg}. After transposing, each row of the register block contains 4 consecutive int8 values from a column of matrix B, which meets the data layout requirement of \texttt{mma}.
Note that each thread has 4 rows in the transposed register block, which are fed into the RHS matrices of 4 \texttt{mma} operations, respectively. Therefore, for two warps within a thread blocks, there are 8 \texttt{mma} operations to be executed in each step, as shown in Figure~\ref{fig:spmmwarps}.

\begin{figure}[ht!]
\centering\includegraphics[width=0.99\linewidth]{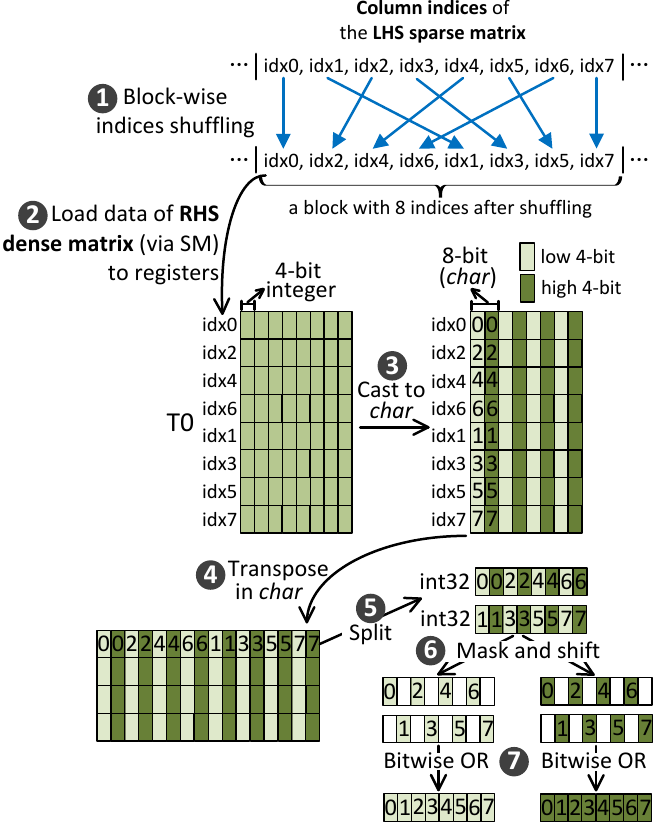}
\caption{\label{fig:indxshuffle} High-performance 4-bit-wise transpose on registers for SpMM by column indices shuffling.}
\end{figure}

Overall, the online transpose strategy for matrix B features coalesced global memory access, conflict-free accesses on shared memory, and high-performance transpose on local registers, which is very efficient as shown in Section~\ref{eval:spmm}.

\subsubsection{Efficient online transpose for matrix B with int4 by indices shuffling}

The online transpose strategy works well for B matrix with int8. However, it may still lead to high overhead for B matrix with int4. After using a similar conflict-free shared memory access approach shown in Figure~\ref{fig:spmmsm}, each thread stores 8*8=64 int4 values on local registers. Since there is no data type in CUDA corresponding to 4-bit integers, transposing these 64 int4 values on local registers results in intensive bit-wise operations. Therefore, we propose a novel column indices shuffling strategy to achieve efficient transpose for int4 values, which is shown in Figure~\ref{fig:indxshuffle}. \one, we modify the SR-BCRS format by shuffling the column indices of the LHS sparse matrix in a block-wise way (block size = 8). The purpose of this shuffling will be recognized in the last step. \two, the corresponding RHS data blocks are loaded into registers similar to that in int8. In \three and \four, the data on local registers are transposed with a granularity of int8 (\texttt{char}). \five, each row of the transposed block is divided into two int32 values. In \six and \seven, after masking, shifting and bitwise OR operations, we get one int32 that only contains the low-4-bit values and another int32 only containing the high-4-bit values of the previous two int32 values. Each int32 has 8 int4 values and the number on each int4 value indicating the ID of the corresponding column index. Amazingly, the corresponding column indices are recovered to the ones before shuffling, and the transpose is also finished. The key benefit in this procedure is that the bitwise operations work only at a granularity of int32, rather than int4. By only using 8 bitwise operations, the transpose of 16 int4 values is finished, which significantly reduces overhead compared with transposing directly with int4 values.

\begin{algorithm}
  \caption{Prefetch the data block of RHS dense matrix}\label{alg:prefetch}
  \begin{algorithmic}[1]
    \Procedure{Pipeline}{$BS_k,nnz$} \Comment{$BS_k$ is the reduction dimension of \texttt{mma}.} \Comment{$nnz$ is the number of nonzero vectors in the row block.}
      \State \_\_$shared$\_\_ double\_buffers\_for\_LHS\_values[];
      \State \_\_$shared$\_\_ LHS\_col\_indices[];
      \State \_\_$registers$\_\_ RHS\_values\_prefetch[];
      \State \_\_$shared$\_\_ RHS\_values[];
      \State $steps$ = $nnz$ / $BS_k$;
      \State \textcolor{blue}{Load\_LHS\_values\_and\_indices\_to\_shared(0)};
      \State \_\_syncthreads();
      \State \textcolor{blue}{Prefetch\_RHS\_values\_to\_registers(0)};
      \For{i=1; i < steps; i++}
            \State \textcolor{blue}{Store\_RHS\_values\_on\_regs\_to\_shared(i-1)};
            \State \textcolor{blue}{Load\_LHS\_values\_and\_indices\_to\_shared(i)};
            \State \_\_syncthreads();
            \State \textcolor{blue}{Prefetch\_RHS\_values\_to\_registers(i)};
            \State \textcolor{blue}{MMA\_compute\_tiles(i-1)};
            \State \_\_syncthreads();
      \EndFor
      \State \textcolor{blue}{Store\_RHS\_values\_on\_regs\_to\_shared(i-1)};
      \State \_\_syncthreads();
      \State \textcolor{blue}{MMA\_compute\_tiles(i-1)};
    \EndProcedure
  \end{algorithmic}
\end{algorithm}

\subsubsection{Prefetch for RHS matrix B}

As discussed in Section~\ref{sec:transpose}, the data block of RHS matrix B are loaded to shared memory for transposing. However, because the matrix A is sparse, there is no data reuse on shared memory for matrix B. Therefore, to mitigate the cost of memory traffic for matrix B, we propose to use a data prefetch strategy, which is shown in Algorithm~\ref{alg:prefetch}. Recall that each thread block requires \textit{nnz}/$BS_{k}$ accumulation steps to get the final results. In addition, there are two phases on the data path when loading data from global memory to shared memory: a) load data from global memory to registers, and b) store data on registers to shared memory. Therefore, we break the load from global to shared into two phases, and compose a pipeline. Lines 7-9 and Line 11 in the first iteration of the \texttt{for} loop are the cold start of the pipeline, after which the first blocks from both A and B are loaded into shared memory. Then, Lines 11-16 in the \texttt{for} loop form the main body of the pipeline. Line 12 loads the LHS dense vectors and the corresponding column indices into shared memory for the next iteration. Using the column indices, the RHS data block is prefetched into registers in Line 14, which is overlapped with the \texttt{mma} computation of the current step in Line 15.
Therefore, the latency of the global memory accesses on RHS blocks is hidden. Note that thread-block-level synchronizations are inserted into the pipeline to achieve thread safety.

\subsection{SDDMM in Magicube}
\begin{figure}[ht!]
\centering\includegraphics[width=0.99\linewidth]{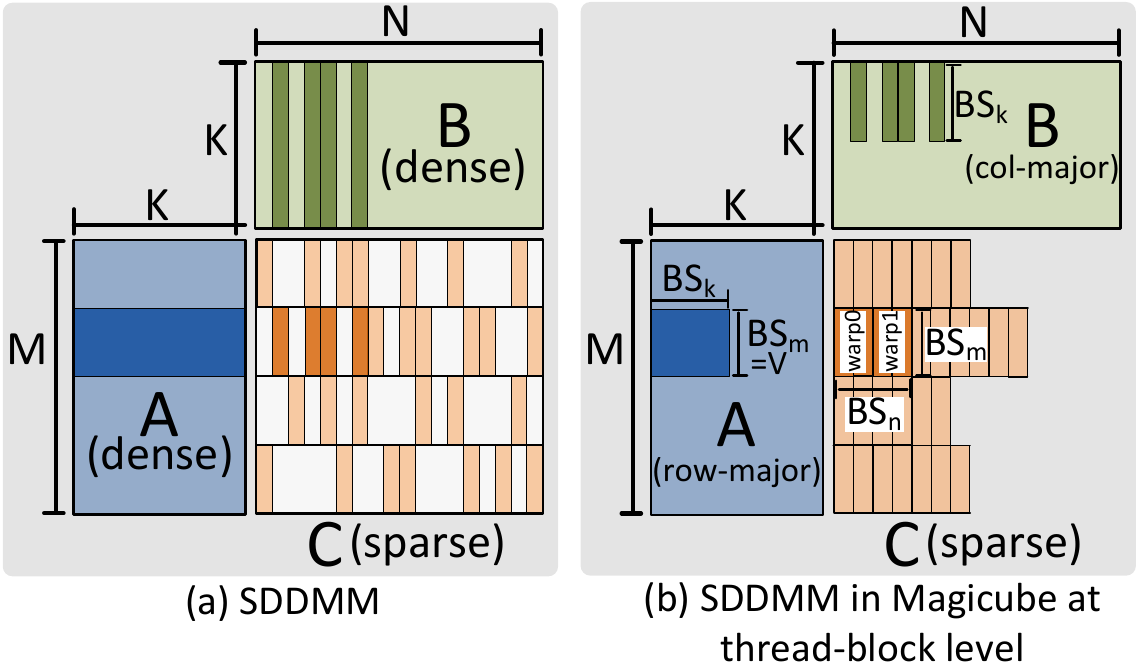}
\caption{\label{fig:sddmmblocks} SDDMM and its thread-block view in Magicube.}
\end{figure}

Sampled dense-dense matrix multiplication (SDDMM) is another major sparse workload in deep learning. The output of SDDMM is a sparse matrix. For example, in the backward pass of a pruned model, the calculation for the sparse weight gradients is an SDDMM operation. In sparse Transformers, the calculation for the sparse attention weights is also an SDDMM operation. Figure~\ref{fig:sddmmblocks}(a) shows an example of SDDMM in which the output matrix C has structured sparsity (i.e., 1-D blocks). The column indices of the dense vectors are used to load the corresponding columns of matrix B.

\begin{figure}[ht!]
\centering\includegraphics[width=0.75\linewidth]{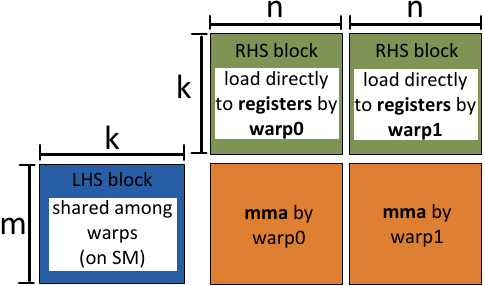}
\caption{\label{fig:sddmmwarps} The warp-level view of the MMAs in SDDMM of Magicube.}
\end{figure}

\subsubsection{Thread blocks of SDDMM}

Figure~\ref{fig:sddmmblocks}(b) shows the implementation of SDDMM in Magicube at the thread-block level. We use row-major storage for A and column-major storage for B. Suppose we have 2 warps in a thread block. Each thread block is responsible for a dense output block of size $BS_m$*$BS_n$. Similar to SpMM, we set $BS_m$=$V$ and $V<=8$. In each step, each thread block calculates the partial results with the reduction dimension $BS_k$. Here $BS_k$ is equal to the reduction dimension of \texttt{mma}. Overall, each thread block needs K/$BS_k$ steps (partial results are accumulated) to obtain the final results.

The LHS block is loaded to shared memory and shared among warps. We also use a similar strategy to Algorithm~\ref{alg:prefetch} to prefetch the LHS data block. Since there is no data reuse for RHS block on shared memory and B is stored in column-major, each thread directly loads the corresponding data into local registers and the data layout requirement of \texttt{mma} is satisfied. The warp-level view of the \texttt{mma} operations is shown in Figure~\ref{fig:sddmmwarps}. Note that the format of the output sparse matrix C is determined by the subsequent operators. If the subsequent operator is SpMM, C is output into SR-BCRS format; if the subsequent operator is softmax, C is output into BCRS format.

\begin{figure}[!ht]
  \centering
    \includegraphics[width=.85\linewidth]{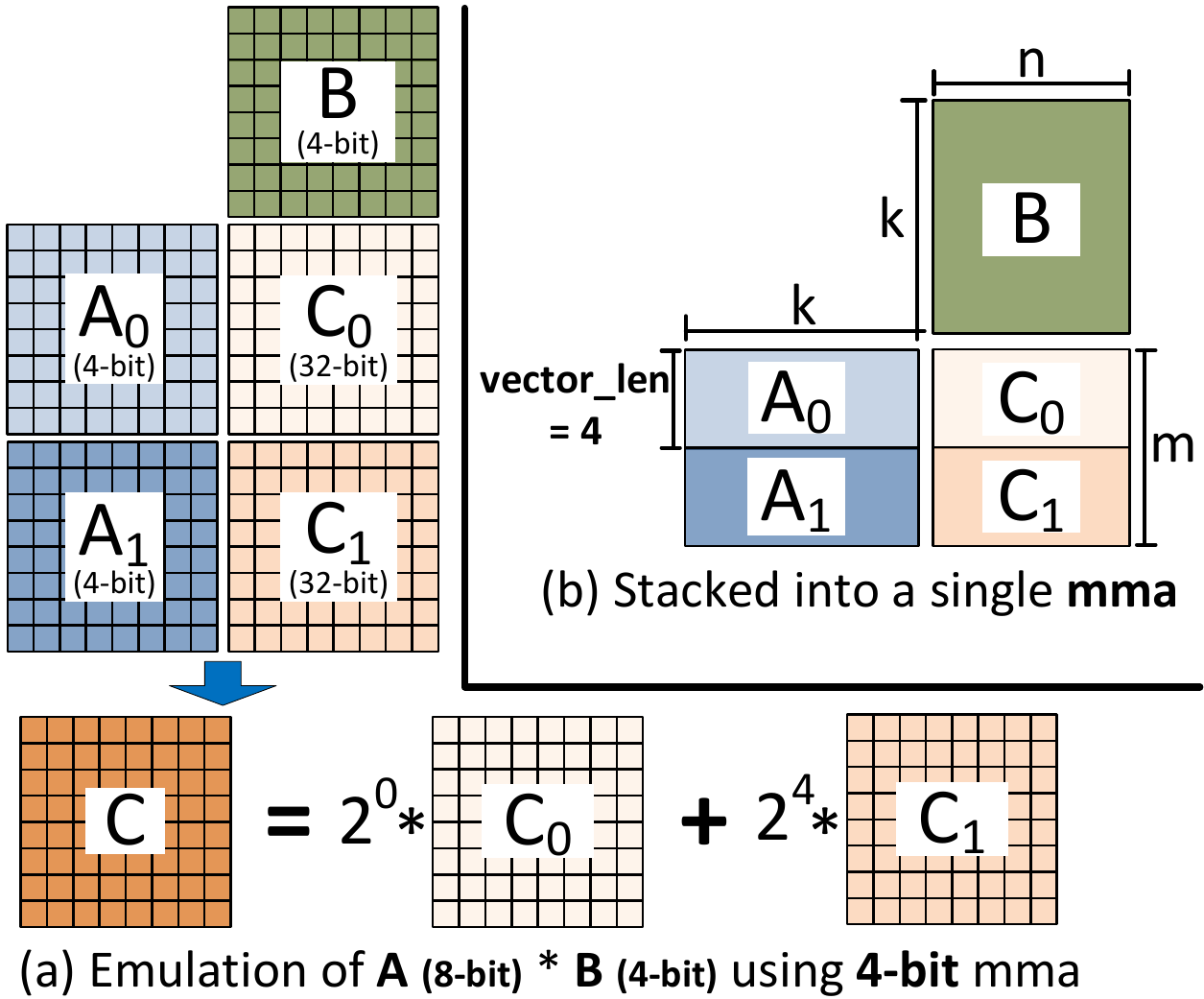}
    \caption{Precision emulation and stacked \texttt{mma}.}
  \label{fig:mmemulate}
\end{figure}

\subsection{Emulation for mixed precision}
\label{sec:emuscheme}
For SpMM, Magicube supports the cases that matrix A has a higher precision than matrix B, for example, A is int8 and B is int4. Quantization schemes with mixed precision~\cite{wang2019haq, zhang2018lq} are well studied in the machine learning community. Since A is already sparse, a higher precision for A may bring higher model accuracy. Magicube also supports that both A and B are in int16 (not natively supported on Tensor cores), for both SpMM and SDDMM. The idea is to divide a high precision value into several low precision values, and then emulate the matrix multiplication mathematically. Arbitrary precision emulation for GEMM (dense) with unsigned integers using 1-bit \texttt{mma} primitives has been studied in~\cite{feng2021apnn}, which requires $x*y$ matrix multiplications in 1-bit integers, where $x$ and $y$ are the number of bits for the values in LHS and RHS matrices, respectively. To tradeoff between mixed precision and low emulation overhead, we only consider precision that the number of bits is a multiple of 4 or 8. Different from previous work, we support precision emulation for both unsigned and signed integers.

\subsubsection{Emulation for unsigned integers}

Here we use the matrix multiplication A (unsigned int8) * B (unsigned int4) as an example. Suppose $a$ = \texttt{11101101} = \texttt{237} (in decimal) is a scalar randomly selected from matrix A. Then, we can directly decompose $a$ to two unsigned int4 values, including $a_{0}$ = \texttt{1101} = 13 (in decimal) for the lower 4 bits and $a_{1}$ = \texttt{1110} = 14 (in decimal) for the higher 4 bits. By mathematical emulation, the original 8-bit values $a$ can be recovered by $a$ = $a_{0}$+$2^4*a_{1}$. Accordingly, the matrix A can be decomposed in to A$_{0}$ containing all the values with lower 4 bits and A$_{1}$ containing all the values with higher 4 bits, as shown in Figure~\ref{fig:mmemulate}(a). Multiplying A$_{0}$ and A$_{1}$ with B generates two intermediate matrices, C$_{0}$ and C$_{1}$. Since the mathematical emulation for scalar $a$ is a linear function, the final output matrix $C$ can be emulated accordingly, i.e., $C$ = $C_{0}$+$2^4*C_{1}$, in which $+$ represents element-wise addition. Note that without mixed precision emulation, matrix B has to be cast into int8, which significantly increases the memory consumption. 

Recall that the vector length $V$ is set to <= $m$. When $V$ < $m$, the Tensor cores are not fully utilized. Here in precision emulation, we find the opportunity to stack multiple partial \texttt{mma} operations into one to increase the utilization of Tensor cores. For example, when $V$=4, we stack two partial \texttt{mma} operations into one to fully utilize Tensor core, as shown in Figure~\ref{fig:mmemulate}(b). The intermediate results C$_{0}$ and C$_{1}$ are distributed among the threads within a warp. Here we use the warp shuffling instruction, i.e., \texttt{\_\_shfl\_xor\_sync}, to exchange intermediate results between threads. At last, the intermediate results are scaled and accumulated to get the final results according to the mathematical emulation. Benefiting from the \texttt{mma} stacking strategy, the utilization of Tensor cores can be increased during precision emulation.

\subsubsection{Emulation for signed integers}

Deep learning models can also be quantized into signed integers using symmetric quantization~\cite{wu2020integer,nagel2021white}. In computers, signed integers are encoded into two's complement. For example, an 8-bit signed integer \texttt{-19} in decimal is encoded to \texttt{11101101}.
We directly decompose it to two 4-bit integers (i.e., \texttt{1110} and \texttt{1101}). However, to guarantee the mathematical correctness, the higher 4 bits must be considered as a signed integer (i.e., \texttt{-2}) and the lower 4 bits must be considered as an unsigned integer (i.e., 13). Then, the mathematical emulation of \texttt{-2*16+13} will recover the two 4-bit integers to the original 8-bit integer (i.e., -19). This means that in the mixed precision emulation for matrix multiplication (Figure~\ref{fig:mmemulate}(a)), the LHS and RHS matrices have different types of integers (i.e., signed and unsigned). Fortunately, \texttt{mma} primitives on Tensor cores support LHS is signed and RHS is unsigned, and vice versa. By considering the highest 4 bits as signed integer and the others as unsigned integer, Magicube supports mixed precision emulation for matrix multiplication with signed integers.

\setlength{\tabcolsep}{3pt}
\begin{table}[!ht]
    \centering
    \caption{The precision supported in Magicube}
    \begin{tabular}{ccccc}
    \toprule
    & Emulated precision & Natively supported \\
    \midrule
    SpMM &  L16-R16, L16-R8, L16-R4, L12-R4, L8-R4 & L8-R8, L4-R4 \\
    SDDMM & L16-R16 & L8-R8, L4-R4\\

    \bottomrule
    \end{tabular}
    \label{tab:precisions}
\end{table}

The supported precision in Magicube is listed in Table~\ref{tab:precisions}, in which L$x$-R$y$ means an $x$-bit LHS matrix multiplied by a $y$-bit RHS matrix.

\begin{figure*}[ht!]
\centering\includegraphics[angle=90, width=0.83\linewidth]{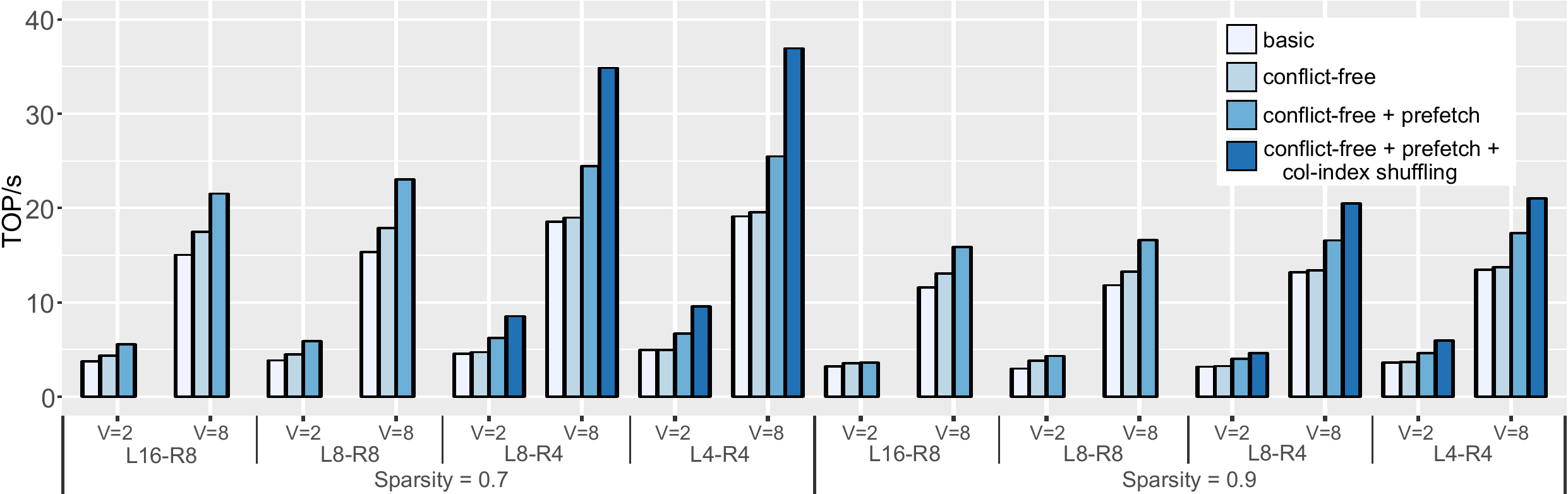}
\caption{\label{fig:spmmopt} Performance evaluation for the optimization methods of SpMM in Magicube with $N$=512. L$x$-R$y$ means $x$-bit LHS matrix and $y$-bit RHS matrix.}
\end{figure*}

\begin{figure*}[ht!]
\centering\includegraphics[angle=90, width=0.95\linewidth]{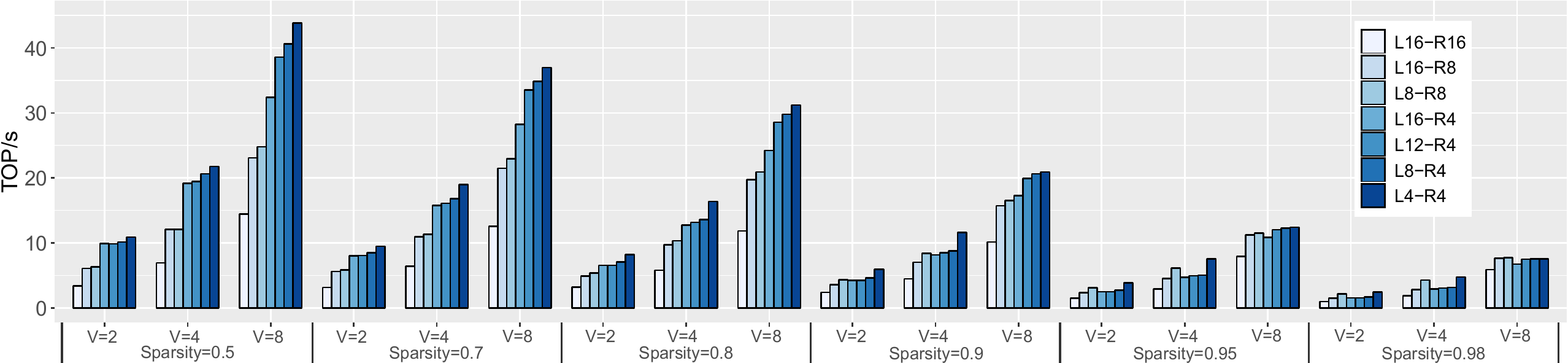}
\caption{\label{fig:spmmpres} The Performance of SpMM in Magicube with different sparsity and precision on A100. L$x$-R$y$ means $x$-bit LHS matrix and $y$-bit RHS matrix.} 
\end{figure*}

\begin{figure*}[ht!]
\centering\includegraphics[angle=90, width=0.95\linewidth]{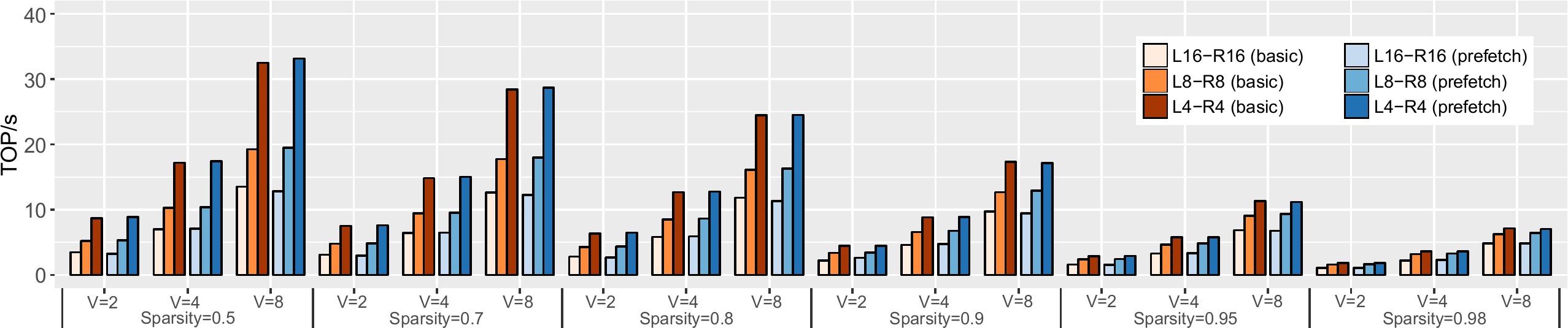}
\caption{\label{fig:sddmmpres} The Performance of SDDMM in Magicube with different sparsity and precision on A100. L$x$-R$y$ means $x$-bit LHS matrix and $y$-bit RHS matrix.} 
\end{figure*}

\section{Evaluation}

\begin{figure*}[ht!]
\centering\includegraphics[angle=90, width=0.98\linewidth]{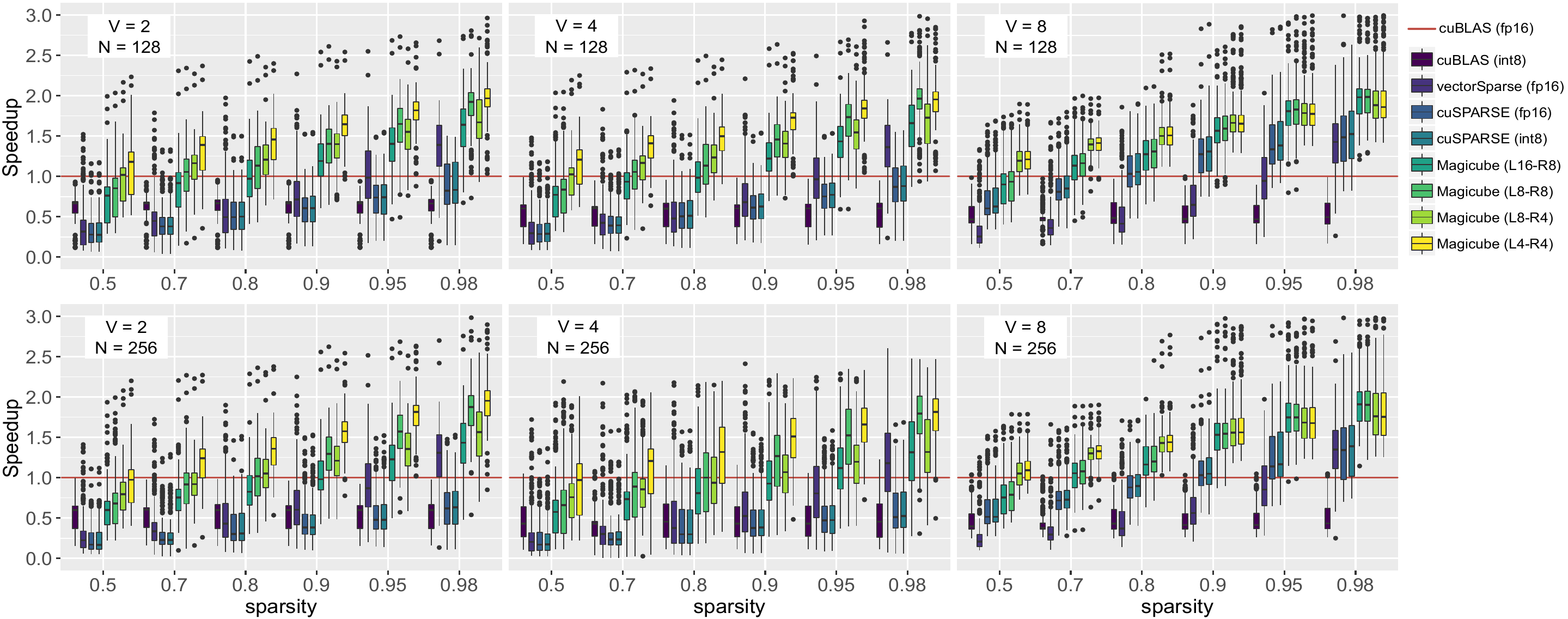}
\caption{\label{fig:spmmbaselines} The Performance of SpMM on A100. The reported speedup is normalized to \textit{cublasHgemm} (dense with fp16). N is the number of columns of the output matrix.} 
\end{figure*}

\begin{figure*}[ht!]
\centering\includegraphics[angle=90, width=0.98\linewidth]{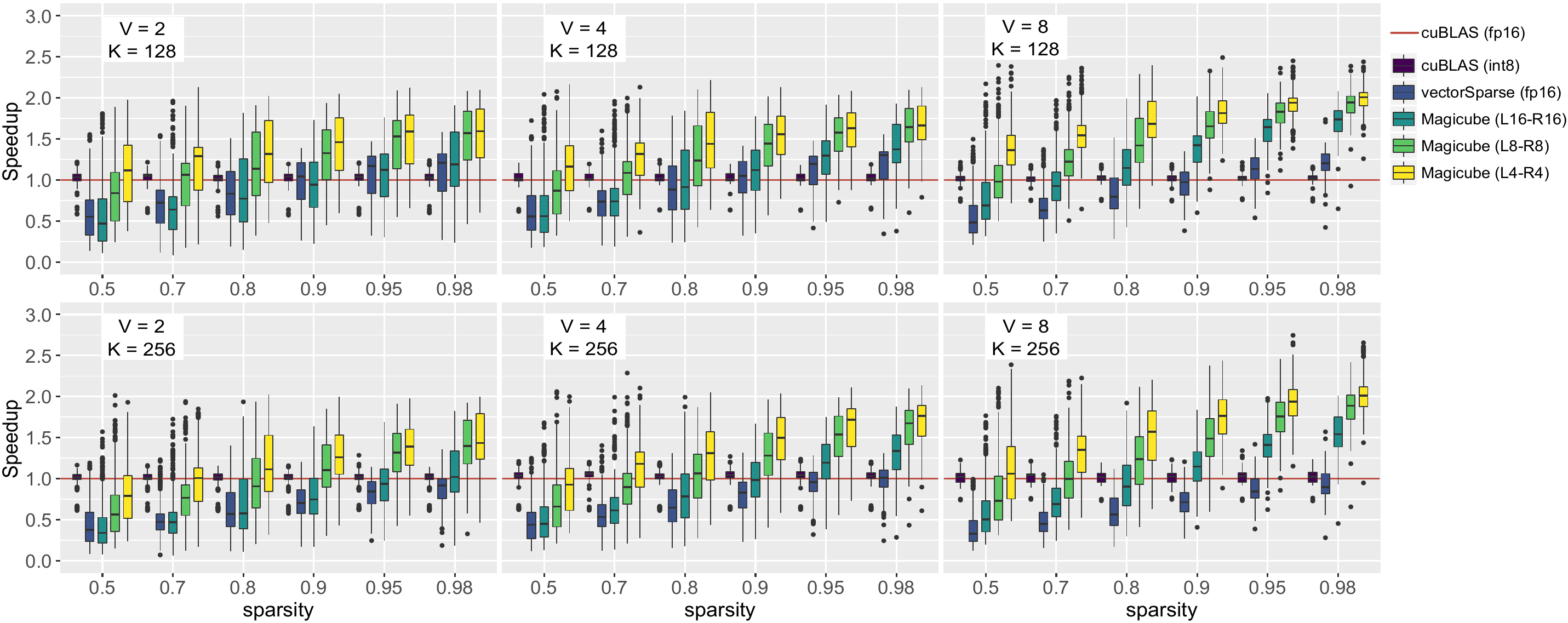}
\caption{\label{fig:sddmmbaselines} The Performance of SDDMM on A100. The reported speedup is normalized to \textit{cublasHgemm} (dense with fp16). K is the reduction dimension.} 
\end{figure*}

We evaluate the performance on an NVIDIA A100 GPU (A100-SXM4-40GB GPU). The A100 GPU has 108 SMs, and each SM has total 192KB configurable L1 cache and shared memory, and 256KB registers. We compare the performance of Magicube with both sparse libraries (cuSPARSE, vectorSparse~\cite{chen2021efficient}) and dense libraries (cuBLAS, cuDNN). We build benchmarks using the Deep Learning Matrix Collection (DLMC)~\cite{DLMC} sparse matrix dataset similar to~\cite{chen2021efficient}, namely a sparse matrix from DLMC is dilated by replacing each scalar with 1-D vectors ($V$ = 2, 4, 8). VectorSparse uses BCRS format (i.e., column vector sparse
encoding). Magicube uses the SR-BCRS format presented in Figure~\ref{fig:sparseformat}. Since cuSPARSE supports SpMM based on Blocked-ELL format, the Blocked-ELL format with the same sparsity and problem size as BCRS and SR-BCRS is generated according to~\cite{chen2021efficient}. In SpMM, the sparse matrices are used for the LHS matrix; In SDDMM, the sparse matrices are used for the output matrix.

We evaluate the performance with different sparsity, which includes 0.5, 0.7, 0.8, 0.9, 0.95, and 0.98. For each sparsity, we select 256 matrices with different sizes from DLMC, which covers all the sparse matrices from ResNet-50 model and part of sparse matrices from Transformer model. Overall, total 1,536 sparse matrices from DLMC are used for evaluation, dilated with different vector length (i.e., $V$ = 2, 4, 8).

In addition to evaluating on micro benchmarks, we also conduct case study on real-world applications - sparse Transformers.

\subsection{Evaluating the optimization strategies in Magicube}

First, we use one sparse matrix (M=256, N=512, K=2304) from DLMC to evaluate the optimization methods proposed for SpMM. The results are shown in Figure~\ref{fig:spmmopt}, in which L$x$-R$y$ means $x$-bit LHS matrix and $y$-bit RHS matrix. With this ablation study, we can see all the optimization methods discussed in Section~\ref{sec:spmm} (including conflict-free shared memory access, data prefetching for RHL data block, and column-index shuffling for 4-bit integers) are very effective. Especially, column-index shuffling proposed for 4-bit integers significantly improve the performance. In the case of L4-R4, V=8 and Sparsity=0.7, the column-index shuffling strategy further improve the performance by 1.45x after all other optimizations are used.

As discussed in Section~\ref{sec:emuscheme}, Magicube supports precision emulation. Figure~\ref{fig:spmmpres} presents the performance of SpMM with mixed precision. On one hand, the main trend is that the lower precision we use, the higher performance we can achieve. But there are several exceptions, for example, L16-R4 has lower performance than L8-R8 when sparsity=0.98, this is because the benefit of memory saving cannot amortize the emulation overhead
when sparsity is high. On the other hand, when the LHS precision is the same, higher precision for RHS matrix does not decrease the performance significantly, which shows the efficiency of the precision emulation strategy in Magicube.

Recall that we also use data prefetch for the LHS data block for SDDMM. But the results in Figure~\ref{fig:sddmmpres} show that prefetching LHS data for SDDMM is not beneficial. This is because the LHS data block in SDDMM is shared and reused among warps. Therefore it already exhibits good performance even without prefetching.

\subsection{Comparison with existing dense and sparse matrix libraries}
\label{eval:spmm}

Next, we compare the performance of Magicube with other dense and sparse libraries. Figure~\ref{fig:spmmbaselines} show the performance of SpMM. We compare Magicube with different precision with cuBLAS (fp16, int8), vectorSparse (fp16), and cuSPARSE (fp16, int8). The speedup shown in the figure is normalized to cuBLAS (fp16). We can see Magicube significantly outperforms all sparse libraries. Magicube can achieve practical speedup over cuBLAS (fp16) with the sparsity higher than 0.7, even with V<8. One interesting point is that cuBLAS (int8) performs even worse than cuBLAS (fp16). In the case of V=8 and N=256, Magicube (L8-R8) outperforms cuSPARSE (int8) by an average (geometric mean) of 1.44x (up to 2.37x), and outperforms cuBLAS (int8, dense) by an average of 2.88x (up to 15.26x) over all the 1,536 matrices; Magicube (L16-R8) outperforms vectorSparse (fp16) by an average of 2.50x (up to 5.27x) over all the 1,536 matrices.

Figure~\ref{fig:sddmmbaselines} show the performance of SDDMM. We compare Magicube with different precision with cuBLAS (fp16, int8) and vectorSparse (fp16). The speedup shown in the figure is normalized to cuBLAS (fp16). Similar to SpMM, SDDMM of Magicube begins to achieve practical speedup over cuBLAS (fp16) with the sparsity higher than 0.7. In the case of V=8 and K=256, Magicube (L16-R16) outperforms vectorSparse (fp16) by an average of 1.58x (up to 2.15x) over all the 1,536 matrices. Higher speedup is achieved by Magicube with lower precision. All these results show the efficiency of quantized sparse kernels in Magicube.

\subsection{Case study with end-to-end sparse Transformer inference}

Transformer models have been widely used in the fields of both natural language processing~\cite{devlin2018bert, brown2020language, zaheer2020big, beltagy2020longformer} and computer vision~\cite{dosovitskiy2020image, yuan2021tokens, liu2021swin, arnab2021vivit}, and exhibit excellent learning ability, thanks to the techniques such as multi-head attention~\cite{vaswani2017attention}. Currently, Transformer is the typical representative of large-scale model workloads. Transformer models commonly have repetitive structures (i.e., the same block repeated multiple times). The network architecture and the model size are mainly determined by the \textbf{head dimension}, the \textbf{number of heads}, and the \textbf{number of layers}. Besides, the workload of self attention grows quadratically with the input \textbf{sequence length}. By varying the parameters above, our evaluation covers different architectures and workloads for Transformers.

We evaluate the performance of Magicube with real-world applications, end-to-end inference with sparse Transformer models from Long-Range Arena (LRA)~\cite{tay2020long}. The model has 4 encoder layers. Sparse Transformer has the workloads of both SDDMM and SpMM in the operation of self attention with a sparse attention mask: 
\begin{align*}
    \mathrm{Attention}(Q,K,V)=\mathrm{softmax}\left(\frac{QK^T\odot M}{\sqrt{d_k}}\right)V\,,
\end{align*}
where $d_k$ is the head dimension ($d_k$ = 64), $M\in\{0,1\}^{L\times L}$ is the sparse attention mask matrix in which $L$ is the sequence length, and $\odot$ represents an element-wise product. For the sparse attention mask matrix, we follow Chen et al.~\cite{chen2021efficient} to add 8x1 vector sparsity constraints.

\begin{figure}[ht!]
\centering\includegraphics[width=0.72\linewidth]{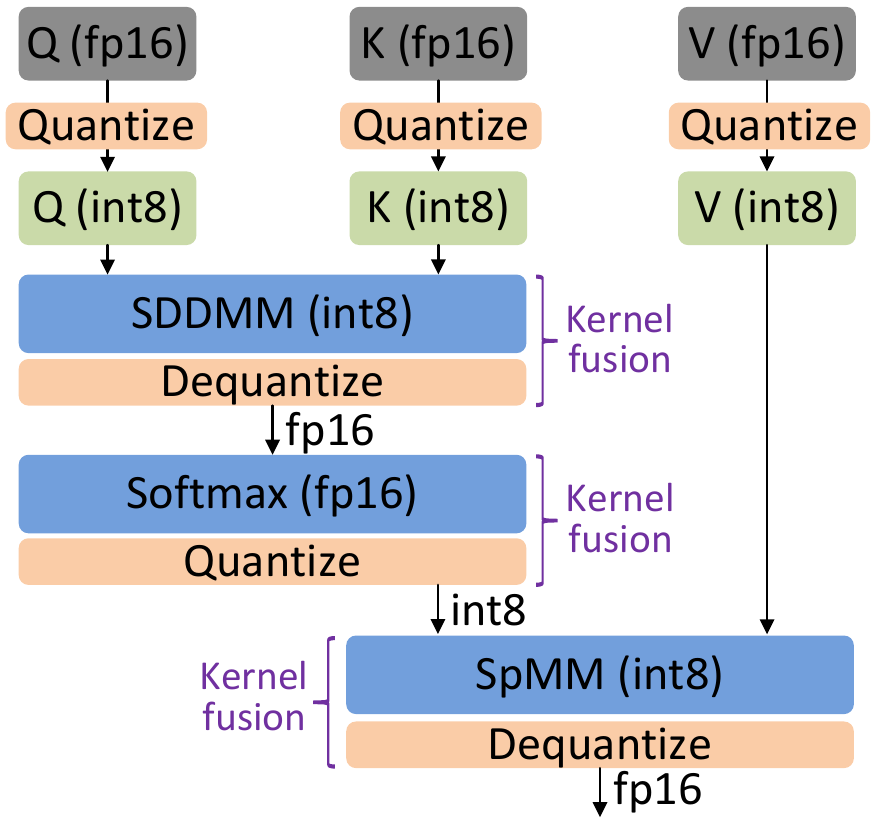}
\caption{\label{fig:sparseTransQ} Quantized self-attention layer with sparse attention mask.} 
\end{figure}

Figure~\ref{fig:sparseTransQ} shows our implementation of a quantized self-attention layer with sparse mask. First, we do quantization for Q, K, V matrices (here Q, K, V are the output of projection). Next, the sparse mask matrix determines the sparsity of the output of $QK^T$. 
Then, $QK^T$ will be a quantized SDDMM operation. We fuse the dequantization with the SDDMM operation. Therefore the output of SDDMM will be in fp16 (half precision). Next, we conduct a softmax kernel in fp16 and fuse quantization with the softmax kernel, and therefore the output of softmax is a sparse matrix with int8. Finally we multiply the int8 sparse attention weight with dense matrix V (int8), which is a SpMM operation. Here we fuse dequantization with SpMM, and the output is a dense matrix in fp16, which will be further input to the following MLP layer.

\begin{figure*}[ht!]
\centering\includegraphics[angle=90, width=0.95\linewidth]{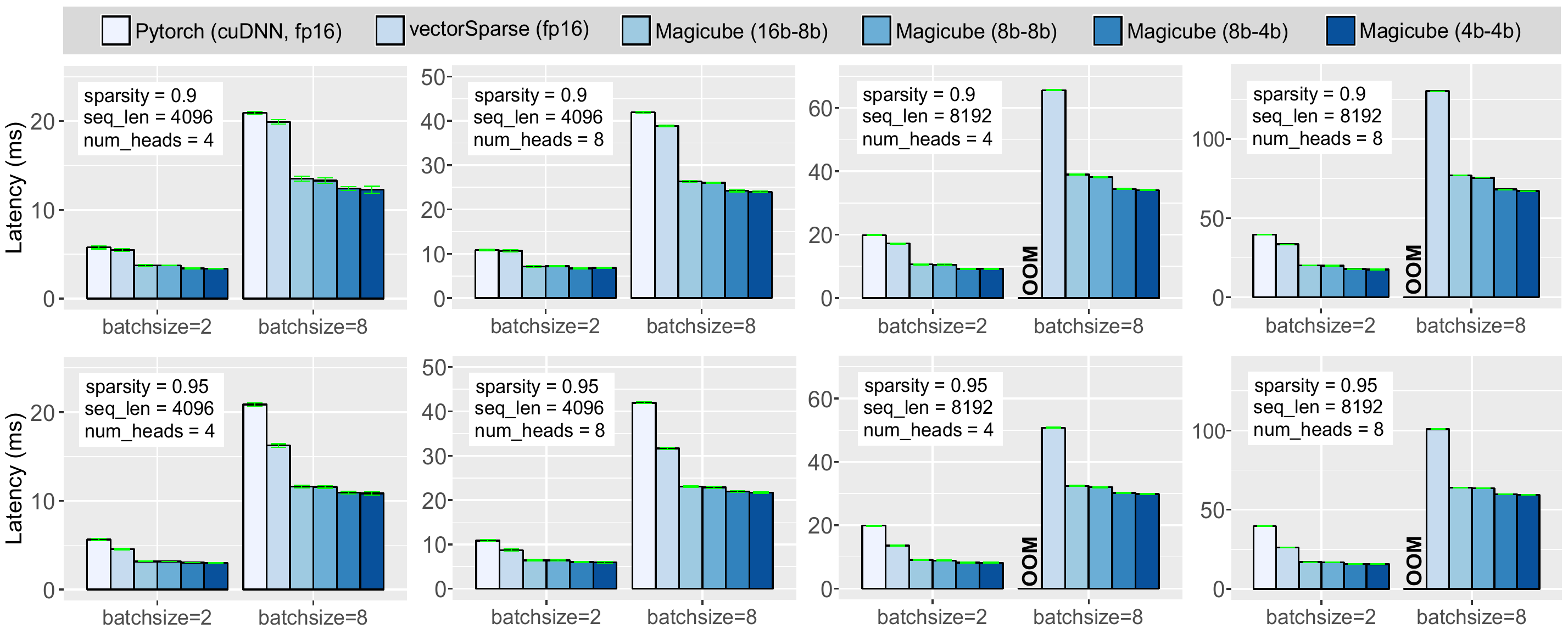}
\caption{\label{fig:sparseTrans} Latency of end-to-end inference of sparse Transformer with different sparsity, sequence length, number of heads, batchsize, and precision.}
\end{figure*}

\begin{table*}[]
\centering
\caption{The test accuracy results of sparse Transformer}
\begin{tabular}{cc|cccccccc}
\toprule
                \multicolumn{2}{c|}{dense} &              \multicolumn{4}{c}{sparsity=0.9} & \multicolumn{4}{c}{sparsity=0.95} \\

                 \cmidrule(lr){1-2} \cmidrule(lr){3-6} \cmidrule(lr){7-10}
                 \begin{tabular}[c]{@{}l@{}}\ \ PyTorch with cuDNN \ \  \\ \ \ \ \ \ \ \ \ \ \ \ (fp32)\ \ \end{tabular} &
                 \begin{tabular}[c]{@{}l@{}}\ \ PyTorch with cuDNN \ \  \\ \ \ \ \ \ \ \ \ \ \ \ (fp16)\ \ \end{tabular} & \begin{tabular}[c]{@{}l@{}}vectorSparse\\ \ \ \ \ (fp16)\end{tabular} 
                 & \begin{tabular}[c]{@{}l@{}}Magicube\\ \ (16b-8b)\end{tabular}
                 & \begin{tabular}[c]{@{}l@{}}Magicube\\ \ \ (8b-8b)\end{tabular} 
                 & \begin{tabular}[c]{@{}l@{}}Magicube\\ \ \ (8b-4b)\end{tabular} 
                 & \begin{tabular}[c]{@{}l@{}}vectorSparse\\ \ \ \ \ (fp16)\end{tabular} 
                 & \begin{tabular}[c]{@{}l@{}}Magicube\\ \ (16b-8b)\end{tabular}
                 & \begin{tabular}[c]{@{}l@{}}Magicube\\ \ \ (8b-8b)\end{tabular} 
                 & \begin{tabular}[c]{@{}l@{}}Magicube\\ \ \ (8b-4b)\end{tabular} \\
                 \midrule
                 57.36\% & \textbf{57.50\%} & 57.14\% & \textbf{57.32\%} & 57.11\% & 56.79\% & \textbf{56.21\%} & 55.79\% & 55.62\% & 55.73\% \\
                 \bottomrule
                 
\end{tabular}
\label{tab:accuracy}
\end{table*}

Figure~\ref{fig:sparseTrans} presents the latency results for end-to-end inference of sparse Transformers with different sparsity, sequence length, number of heads, batchsize, and precision. We keep the number of encoder layers to 4 since the total workload is proportional to the number of layers, and keep head dimension to 64 which is commonly used in Transformer models. For Magicube, $x$b-$y$b means quantizing the output of softmax to $x$-bit integers and quantizing Q, K, V to $y$-bit integers. We compare our implementation with the fp16 dense counterpart (PyTorch 1.9 with cuDNN version 8005) and vectorSparse (fp16). We run each setup for 256 iterations, and the green bars on the histogram indicates the 95\% confidence interval. For end-to-end inference with sparsity=90\%, seq\_len=4,096, and num\_heads=4, Magicube achieves 1.43x-1.63x speedup over vectorSparse (the state-of-the-art sparse library with fp16 on Tensor cores), and achieves 1.50x-1.70x over PyTorch with cuDNN (dense). After increasing the sequence length to 8,192, the fp16 dense counterpart runs out of memory when batchsize=8, since the memory cost of self attention grows quadratically with the sequence length. When sparsity=90\%, seq\_len=8,192, and num\_heads=4, Magicube achieves 1.62x-1.92x speedup over vectorSparse, which indicates that Magicube enables higher speedups for longer sequences. By increasing num\_heads from 4 to 8, the runtime for all schemes increases by about 2x. By increasing sparsity from 90\% to 95\%, the sparse libraries (vectorSparse and Magicube) further reduce the latency compared to the dense counterpart.

\newpage

Table~V presents the test accuracy results for text classification using sparse Transformer with num\_heads=4 and seq\_len=4,096. Since LRA's repository uses an outdated version of FLAX and we cannot make it run, we reimplemented it in PyTorch. We train the model with dense and sparse attention masks using the same hyperparameters, and finetune it for quantization. Compared with the PyTorch with cuDNN (dense) and sparse (sparsity=90\%) models with fp16, the sparse (sparsity=90\%) model with quantization (16-bit softmax output and 8-bit Q, K, V) achieves a comparable accuracy. By reducing the precision of softmax output to 8 bits, the accuracy drops slightly, which indicates that a higher precision of softmax helps to maintain the accuracy. After further increasing the sparsity to 95\%, the accuracy of sparse models drops within acceptable ranges.

\section{Discussion}

\paragraph{Generality to other AI accelerators} Although Magicube is designed on NVIDIA GPUs with Tensor cores, the key insights of Magicube can be easily adapted to other AI accelerators. For example, AMD MI250X GPU~\cite{MI200} provides 383.0 TOP/s peak performance for int8 through the technology of Matrix Core. To program on Matrix cores, AMD provides wavefront-level instructions with the semantic of Matrix Fused Multiply-Add (\texttt{MFMA}). Similar to \texttt{mma} on NVIDIA GPUs, \texttt{MFMA} instructions (e.g., V\_MFMA\_I32\_16X16X16I8) also have specific requirements for the data layout. The techniques of Magicube, such as the SR-BCRS format, online transpose, and data prefetching, can also be utilized on such accelerators.

\paragraph{Work with distributed deep learning systems} Efficient large-scale deep learning on distributed systems is commonly realized by a combination of data, operator, and pipeline parallelism~\cite{megatron, li2021chimera, shoeybi2019megatron, zheng2022alpa}. Magicube can be used in these distributed deep learning systems as the backend compute library to accelerate each (sub-)operator and alleviate the computation bottleneck. How to maintain the convergence quality while introducing sparsity and quantization is our future work.

\paragraph{More applications} In  this work, We focus on large-scale models based on Transformer, in which the sparse workload is introduced by the sparse attention mask. Our scheme may also benefit other sparse workloads in deep learning. For example, training with model pruning results in SpMM in the forward pass and SDDMM in the backward pass. Furthermore, our empirical observation shows that the curvature matrices in second-order optimization~\cite{pauloski2021kaisa, osawa2019large} may also be approximated through sparsity.

\section{Conclusion}
\label{sec:conclusion}
In this paper, we propose \textit{Magicube}, a high-performance sparse-matrix library with low-precision (8-bit, 4-bit) integers on Tensor cores for SpMM and SDDMM to accelerate sparse and quantized matrix operations in deep learning. 
For compressing the quantities with low-precision integers, we design a Strided Row-major BCRS (SR-BCRS) format. 
Through the performance evaluation with over 1,536 sparse matrices with different sizes and sparisy (50-98\%) from DLMC \cite{DLMC} dataset on an NVIDIA A100 GPU, we demonstrate that Magicube achieves on average 1.44x (up to 2.37x) speedup for SpMM over cuSPARSE.
We also demonstrate that for end-to-end inference of a Transformer \cite{vaswani2017attention} model with a sparse (90\%) attention map, Magicube achieves 1.43x speedup over vectorSparse~\cite{chen2021efficient} (the state-of-the-art sparse library with fp16) and 1.50x speedup over PyTorch with cuDNN (the fp16 dense library), with a comparable accuracy. The only constraint our sparse kernels impose on the data layout is that each nonzero block is 1-D blocks of shape e.g., $8\times1$, $4\times 1$, $2\times 1$.
Under this constraint, Magicube successfully gains practical speedups from both the sparsity and quantization in deep learning workloads. 
We foresee this work will motivate future research on sparsification and quantization methods for deep learning models that lead to more effective utilization of modern accelerators.

\section*{Acknowledgment}
This work receives EuroHPC-JU funding under program EUPilot, grant No. 101034126, with support from the Horizon2020 programme. K.O. is supported by the ETH Postdoctoral Fellowship. We also thank the Swiss National Supercomputing Center (CSCS) for providing computing resources.

\bibliographystyle{IEEEtran}
\bibliography{IEEEabrv,mybib}

\begin{thebibliography}{10}
\providecommand{\url}[1]{#1}
\csname url@samestyle\endcsname
\providecommand{\newblock}{\relax}
\providecommand{\bibinfo}[2]{#2}
\providecommand{\BIBentrySTDinterwordspacing}{\spaceskip=0pt\relax}
\providecommand{\BIBentryALTinterwordstretchfactor}{4}
\providecommand{\BIBentryALTinterwordspacing}{\spaceskip=\fontdimen2\font plus
\BIBentryALTinterwordstretchfactor\fontdimen3\font minus
  \fontdimen4\font\relax}
\providecommand{\BIBforeignlanguage}[2]{{%
\expandafter\ifx\csname l@#1\endcsname\relax
\typeout{** WARNING: IEEEtran.bst: No hyphenation pattern has been}%
\typeout{** loaded for the language `#1'. Using the pattern for}%
\typeout{** the default language instead.}%
\else
\language=\csname l@#1\endcsname
\fi
#2}}
\providecommand{\BIBdecl}{\relax}
\BIBdecl

\bibitem{hestness2017deep}
J.~Hestness, S.~Narang, N.~Ardalani, G.~Diamos, H.~Jun, H.~Kianinejad,
  M.~Patwary, M.~Ali, Y.~Yang, and Y.~Zhou, ``Deep learning scaling is
  predictable, empirically,'' \emph{arXiv preprint arXiv:1712.00409}, 2017.

\bibitem{rosenfeld2019constructive}
J.~S. Rosenfeld, A.~Rosenfeld, Y.~Belinkov, and N.~Shavit, ``A constructive
  prediction of the generalization error across scales,'' \emph{arXiv preprint
  arXiv:1909.12673}, 2019.

\bibitem{kaplan2020scaling}
J.~Kaplan, S.~McCandlish, T.~Henighan, T.~B. Brown, B.~Chess, R.~Child,
  S.~Gray, A.~Radford, J.~Wu, and D.~Amodei, ``Scaling laws for neural language
  models,'' \emph{arXiv preprint arXiv:2001.08361}, 2020.

\bibitem{brown2020language}
T.~Brown, B.~Mann, N.~Ryder, M.~Subbiah, J.~D. Kaplan, P.~Dhariwal,
  A.~Neelakantan, P.~Shyam, G.~Sastry, A.~Askell \emph{et~al.}, ``Language
  models are few-shot learners,'' \emph{Advances in neural information
  processing systems}, vol.~33, pp. 1877--1901, 2020.

\bibitem{openai2018aiandcompute}
\BIBentryALTinterwordspacing
{OpenAI}, ``{AI} and compute,'' 2018. [Online]. Available:
  \url{https://openai.com/blog/ai-and-compute/}
\BIBentrySTDinterwordspacing

\bibitem{strubell2019energy}
E.~Strubell, A.~Ganesh, and A.~McCallum, ``Energy and policy considerations for
  deep learning in {NLP},'' in \emph{Proceedings of the 57th Annual Meeting of
  the Association for Computational Linguistics (ACL)}, 2019.

\bibitem{patterson2021carbon}
D.~Patterson, J.~Gonzalez, Q.~Le, C.~Liang, L.-M. Munguia, D.~Rothchild, D.~So,
  M.~Texier, and J.~Dean, ``Carbon emissions and large neural network
  training,'' \emph{arXiv preprint arXiv:2104.10350}, 2021.

\bibitem{gholami2021survey}
A.~Gholami, S.~Kim, Z.~Dong, Z.~Yao, M.~W. Mahoney, and K.~Keutzer, ``A survey
  of quantization methods for efficient neural network inference,'' \emph{arXiv
  preprint arXiv:2103.13630}, 2021.

\bibitem{hoefler2021sparsity}
T.~Hoefler, D.~Alistarh, T.~Ben-Nun, N.~Dryden, and A.~Peste, ``Sparsity in
  deep learning: Pruning and growth for efficient inference and training in
  neural networks,'' \emph{Journal of Machine Learning Research}, vol.~22, no.
  241, pp. 1--124, 2021.

\bibitem{cusparse}
{NVIDIA}, ``{cuSPARSE},'' https://developer.nvidia.com/cusparse.

\bibitem{cusparselt}
------, ``{Exploiting NVIDIA Ampere Structured Sparsity with cuSPARSELt },''
  December 2020,
  https://developer.nvidia.com/blog/exploiting-ampere-structured-sparsity-with-cusparselt/.

\bibitem{SparseTC}
J.~Pool, A.~Sawarkar, and J.~Rodge, ``Accelerating inference with sparsity
  using the nvidia ampere architecture and nvidia tensorrt,'' 2021.

\bibitem{gale2020sparse}
T.~Gale, M.~Zaharia, C.~Young, and E.~Elsen, ``Sparse gpu kernels for deep
  learning,'' in \emph{SC20: International Conference for High Performance
  Computing, Networking, Storage and Analysis}.\hskip 1em plus 0.5em minus
  0.4em\relax IEEE, 2020, pp. 1--14.

\bibitem{chen2021efficient}
Z.~Chen, Z.~Qu, L.~Liu, Y.~Ding, and Y.~Xie, ``Efficient tensor core-based gpu
  kernels for structured sparsity under reduced precision,'' in
  \emph{Proceedings of the International Conference for High Performance
  Computing, Networking, Storage and Analysis}, 2021, pp. 1--14.

\bibitem{van2020bayesian}
M.~Van~Baalen, C.~Louizos, M.~Nagel, R.~A. Amjad, Y.~Wang, T.~Blankevoort, and
  M.~Welling, ``Bayesian bits: Unifying quantization and pruning,''
  \emph{Advances in neural information processing systems}, vol.~33, pp.
  5741--5752, 2020.

\bibitem{nvidia2017nvidia}
NVIDIA, ``Nvidia tesla v100 gpu architecture,'' 2017.

\bibitem{jia2018dissecting}
Z.~Jia, M.~Maggioni, B.~Staiger, and D.~P. Scarpazza, ``Dissecting the nvidia
  volta gpu architecture via microbenchmarking,'' \emph{arXiv preprint
  arXiv:1804.06826}, 2018.

\bibitem{amperetuning}
{NVIDIA}, ``{Tuning CUDA Applications for Ampere},'' September 2021,
  https://docs.nvidia.com/cuda/ampere-tuning-guide/index.html.

\bibitem{child2019generating}
R.~Child, S.~Gray, A.~Radford, and I.~Sutskever, ``Generating long sequences
  with sparse transformers,'' \emph{arXiv preprint arXiv:1904.10509}, 2019.

\bibitem{tung2018deep}
F.~Tung and G.~Mori, ``Deep neural network compression by in-parallel
  pruning-quantization,'' \emph{IEEE transactions on pattern analysis and
  machine intelligence}, vol.~42, no.~3, pp. 568--579, 2018.

\bibitem{yang2020automatic}
H.~Yang, S.~Gui, Y.~Zhu, and J.~Liu, ``Automatic neural network compression by
  sparsity-quantization joint learning: A constrained optimization-based
  approach,'' in \emph{Proceedings of the IEEE/CVF Conference on Computer
  Vision and Pattern Recognition}, 2020, pp. 2178--2188.

\bibitem{yu2020joint}
P.-H. Yu, S.-S. Wu, J.~P. Klopp, L.-G. Chen, and S.-Y. Chien, ``Joint pruning
  \& quantization for extremely sparse neural networks,'' \emph{arXiv preprint
  arXiv:2010.01892}, 2020.

\bibitem{wang2020apq}
T.~Wang, K.~Wang, H.~Cai, J.~Lin, Z.~Liu, H.~Wang, Y.~Lin, and S.~Han, ``Apq:
  Joint search for network architecture, pruning and quantization policy,'' in
  \emph{Proceedings of the IEEE/CVF Conference on Computer Vision and Pattern
  Recognition}, 2020, pp. 2078--2087.

\bibitem{srivastava2019joint}
G.~Srivastava, D.~Kadetotad, S.~Yin, V.~Berisha, C.~Chakrabarti, and J.-s. Seo,
  ``Joint optimization of quantization and structured sparsity for compressed
  deep neural networks,'' in \emph{ICASSP 2019-2019 IEEE International
  Conference on Acoustics, Speech and Signal Processing (ICASSP)}.\hskip 1em
  plus 0.5em minus 0.4em\relax IEEE, 2019, pp. 1393--1397.

\bibitem{han2015deep}
S.~Han, H.~Mao, and W.~J. Dally, ``Deep compression: Compressing deep neural
  networks with pruning, trained quantization and huffman coding,'' \emph{arXiv
  preprint arXiv:1510.00149}, 2015.

\bibitem{you2019drawing}
H.~You, C.~Li, P.~Xu, Y.~Fu, Y.~Wang, X.~Chen, R.~G. Baraniuk, Z.~Wang, and
  Y.~Lin, ``Drawing early-bird tickets: Towards more efficient training of deep
  networks,'' \emph{arXiv preprint arXiv:1909.11957}, 2019.

\bibitem{rastegari2016xnor}
M.~Rastegari, V.~Ordonez, J.~Redmon, and A.~Farhadi, ``Xnor-net: Imagenet
  classification using binary convolutional neural networks,'' in
  \emph{European conference on computer vision}.\hskip 1em plus 0.5em minus
  0.4em\relax Springer, 2016, pp. 525--542.

\bibitem{micikevicius2017mixed}
P.~Micikevicius, S.~Narang, J.~Alben, G.~Diamos, E.~Elsen, D.~Garcia,
  B.~Ginsburg, M.~Houston, O.~Kuchaiev, G.~Venkatesh \emph{et~al.}, ``Mixed
  precision training,'' \emph{arXiv preprint arXiv:1710.03740}, 2017.

\bibitem{wu2020integer}
H.~Wu, P.~Judd, X.~Zhang, M.~Isaev, and P.~Micikevicius, ``Integer quantization
  for deep learning inference: Principles and empirical evaluation,''
  \emph{arXiv preprint arXiv:2004.09602}, 2020.

\bibitem{hubara2021accurate}
I.~Hubara, Y.~Nahshan, Y.~Hanani, R.~Banner, and D.~Soudry, ``Accurate post
  training quantization with small calibration sets,'' in \emph{International
  Conference on Machine Learning}.\hskip 1em plus 0.5em minus 0.4em\relax PMLR,
  2021, pp. 4466--4475.

\bibitem{nagel2021white}
M.~Nagel, M.~Fournarakis, R.~A. Amjad, Y.~Bondarenko, M.~van Baalen, and
  T.~Blankevoort, ``A white paper on neural network quantization,'' \emph{arXiv
  preprint arXiv:2106.08295}, 2021.

\bibitem{vaswani2017attention}
A.~Vaswani, N.~Shazeer, N.~Parmar, J.~Uszkoreit, L.~Jones, A.~N. Gomez,
  {\L}.~Kaiser, and I.~Polosukhin, ``Attention is all you need,''
  \emph{Advances in neural information processing systems}, vol.~30, 2017.

\bibitem{bahdanau2014neural}
D.~Bahdanau, K.~Cho, and Y.~Bengio, ``Neural machine translation by jointly
  learning to align and translate,'' \emph{arXiv preprint arXiv:1409.0473},
  2014.

\bibitem{devlin2018bert}
J.~Devlin, M.-W. Chang, K.~Lee, and K.~Toutanova, ``Bert: Pre-training of deep
  bidirectional transformers for language understanding,'' \emph{arXiv preprint
  arXiv:1810.04805}, 2018.

\bibitem{sanh2020movement}
V.~Sanh, T.~Wolf, and A.~Rush, ``Movement pruning: Adaptive sparsity by
  fine-tuning,'' \emph{Advances in Neural Information Processing Systems},
  vol.~33, pp. 20\,378--20\,389, 2020.

\bibitem{lagunas2021block}
F.~Lagunas, E.~Charlaix, V.~Sanh, and A.~M. Rush, ``Block pruning for faster
  transformers,'' \emph{arXiv preprint arXiv:2109.04838}, 2021.

\bibitem{mao2021tprune}
J.~Mao, H.~Yang, A.~Li, H.~Li, and Y.~Chen, ``Tprune: Efficient transformer
  pruning for mobile devices,'' \emph{ACM Transactions on Cyber-Physical
  Systems}, vol.~5, no.~3, pp. 1--22, 2021.

\bibitem{bhandare2019efficient}
A.~Bhandare, V.~Sripathi, D.~Karkada, V.~Menon, S.~Choi, K.~Datta, and
  V.~Saletore, ``Efficient 8-bit quantization of transformer neural machine
  language translation model,'' \emph{arXiv preprint arXiv:1906.00532}, 2019.

\bibitem{zafrir2019q8bert}
O.~Zafrir, G.~Boudoukh, P.~Izsak, and M.~Wasserblat, ``Q8bert: Quantized 8bit
  bert,'' in \emph{2019 Fifth Workshop on Energy Efficient Machine Learning and
  Cognitive Computing-NeurIPS Edition (EMC2-NIPS)}.\hskip 1em plus 0.5em minus
  0.4em\relax IEEE, 2019, pp. 36--39.

\bibitem{mao2020ladabert}
Y.~Mao, Y.~Wang, C.~Wu, C.~Zhang, Y.~Wang, Y.~Yang, Q.~Zhang, Y.~Tong, and
  J.~Bai, ``Ladabert: Lightweight adaptation of bert through hybrid model
  compression,'' \emph{arXiv preprint arXiv:2004.04124}, 2020.

\bibitem{kim2021bert}
S.~Kim, A.~Gholami, Z.~Yao, M.~W. Mahoney, and K.~Keutzer, ``I-bert:
  Integer-only bert quantization,'' in \emph{International conference on
  machine learning}.\hskip 1em plus 0.5em minus 0.4em\relax PMLR, 2021, pp.
  5506--5518.

\bibitem{wang2020linformer}
S.~Wang, B.~Z. Li, M.~Khabsa, H.~Fang, and H.~Ma, ``Linformer: Self-attention
  with linear complexity,'' \emph{arXiv preprint arXiv:2006.04768}, 2020.

\bibitem{zaheer2020big}
M.~Zaheer, G.~Guruganesh, K.~A. Dubey, J.~Ainslie, C.~Alberti, S.~Ontanon,
  P.~Pham, A.~Ravula, Q.~Wang, L.~Yang \emph{et~al.}, ``Big bird: Transformers
  for longer sequences,'' \emph{Advances in Neural Information Processing
  Systems}, vol.~33, pp. 17\,283--17\,297, 2020.

\bibitem{beltagy2020longformer}
I.~Beltagy, M.~E. Peters, and A.~Cohan, ``Longformer: The long-document
  transformer,'' \emph{arXiv preprint arXiv:2004.05150}, 2020.

\bibitem{choromanski2020rethinking}
K.~Choromanski, V.~Likhosherstov, D.~Dohan, X.~Song, A.~Gane, T.~Sarlos,
  P.~Hawkins, J.~Davis, A.~Mohiuddin, L.~Kaiser \emph{et~al.}, ``Rethinking
  attention with performers,'' \emph{arXiv preprint arXiv:2009.14794}, 2020.

\bibitem{han2016eie}
S.~Han, X.~Liu, H.~Mao, J.~Pu, A.~Pedram, M.~A. Horowitz, and W.~J. Dally,
  ``Eie: Efficient inference engine on compressed deep neural network,''
  \emph{ACM SIGARCH Computer Architecture News}, vol.~44, no.~3, pp. 243--254,
  2016.

\bibitem{li2019bstc}
A.~Li, T.~Geng, T.~Wang, M.~Herbordt, S.~L. Song, and K.~Barker, ``Bstc: A
  novel binarized-soft-tensor-core design for accelerating bit-based
  approximated neural nets,'' in \emph{Proceedings of the International
  Conference for High Performance Computing, Networking, Storage and Analysis},
  2019, pp. 1--30.

\bibitem{li2020accelerating}
A.~Li and S.~Su, ``Accelerating binarized neural networks via bit-tensor-cores
  in turing gpus,'' \emph{IEEE Transactions on Parallel and Distributed
  Systems}, vol.~32, no.~7, pp. 1878--1891, 2020.

\bibitem{feng2021apnn}
B.~Feng, Y.~Wang, T.~Geng, A.~Li, and Y.~Ding, ``Apnn-tc: Accelerating
  arbitrary precision neural networks on ampere gpu tensor cores,'' in
  \emph{Proceedings of the International Conference for High Performance
  Computing, Networking, Storage and Analysis}, 2021, pp. 1--13.

\bibitem{wang2018swsptrsv}
X.~Wang, W.~Liu, W.~Xue, and L.~Wu, ``swsptrsv: a fast sparse triangular solve
  with sparse level tile layout on sunway architectures,'' in \emph{Proceedings
  of the 23rd ACM SIGPLAN Symposium on Principles and Practice of Parallel
  Programming}, 2018, pp. 338--353.

\bibitem{chen2018performance}
Y.~Chen, K.~Li, W.~Yang, G.~Xiao, X.~Xie, and T.~Li, ``Performance-aware model
  for sparse matrix-matrix multiplication on the sunway taihulight
  supercomputer,'' \emph{IEEE transactions on parallel and distributed
  systems}, vol.~30, no.~4, pp. 923--938, 2018.

\bibitem{xie2021fast}
C.~Xie, J.~Chen, J.~Firoz, J.~Li, S.~L. Song, K.~Barker, M.~Raugas, and A.~Li,
  ``Fast and scalable sparse triangular solver for multi-gpu based hpc
  architectures,'' in \emph{50th International Conference on Parallel
  Processing}, 2021, pp. 1--11.

\bibitem{xie2019ia}
Z.~Xie, G.~Tan, W.~Liu, and N.~Sun, ``Ia-spgemm: An input-aware auto-tuning
  framework for parallel sparse matrix-matrix multiplication,'' in
  \emph{Proceedings of the ACM International Conference on Supercomputing},
  2019, pp. 94--105.

\bibitem{niu2021tilespmv}
Y.~Niu, Z.~Lu, M.~Dong, Z.~Jin, W.~Liu, and G.~Tan, ``Tilespmv: A tiled
  algorithm for sparse matrix-vector multiplication on gpus,'' in \emph{2021
  IEEE International Parallel and Distributed Processing Symposium
  (IPDPS)}.\hskip 1em plus 0.5em minus 0.4em\relax IEEE, 2021, pp. 68--78.

\bibitem{xie2021pattern}
Z.~Xie, G.~Tan, W.~Liu, and N.~Sun, ``A pattern-based spgemm library for
  multi-core and many-core architectures,'' \emph{IEEE Transactions on Parallel
  and Distributed Systems}, vol.~33, no.~1, pp. 159--175, 2021.

\bibitem{nvptx}
NVIDIA, ``Parallel thread execution isa application guide,'' 2021.

\bibitem{nvhopper}
\BIBentryALTinterwordspacing
{Michael Andersch, Greg Palmer, Ronny Krashinsky, Nick Stam, Vishal Mehta,
  Gonzalo Brito and Sridhar Ramaswamy}, ``{NVIDIA Hopper Architecture
  In-Depth},'' March 2022. [Online]. Available:
  \url{https://developer.nvidia.com/blog/nvidia-hopper-architecture-in-depth/}
\BIBentrySTDinterwordspacing

\bibitem{mao2017exploring}
H.~Mao, S.~Han, J.~Pool, W.~Li, X.~Liu, Y.~Wang, and W.~J. Dally, ``Exploring
  the regularity of sparse structure in convolutional neural networks,''
  \emph{arXiv preprint arXiv:1705.08922}, 2017.

\bibitem{vuduc2005oski}
R.~Vuduc, J.~W. Demmel, and K.~A. Yelick, ``Oski: A library of automatically
  tuned sparse matrix kernels,'' in \emph{Journal of Physics: Conference
  Series}, vol.~16, no.~1.\hskip 1em plus 0.5em minus 0.4em\relax IOP
  Publishing, 2005, p. 071.

\bibitem{liu2021sparta}
J.~Liu, J.~Ren, R.~Gioiosa, D.~Li, and J.~Li, ``Sparta: High-performance,
  element-wise sparse tensor contraction on heterogeneous memory,'' in
  \emph{Proceedings of the 26th ACM SIGPLAN Symposium on Principles and
  Practice of Parallel Programming}, 2021, pp. 318--333.

\bibitem{pinar1999improving}
A.~Pinar and M.~T. Heath, ``Improving performance of sparse matrix-vector
  multiplication,'' in \emph{SC'99: Proceedings of the 1999 ACM/IEEE Conference
  on Supercomputing}.\hskip 1em plus 0.5em minus 0.4em\relax IEEE, 1999, pp.
  30--30.

\bibitem{cudaguide}
{NVIDIA}, ``{CUDA C++ Programming Guide},'' September 2021.

\bibitem{wang2019haq}
K.~Wang, Z.~Liu, Y.~Lin, J.~Lin, and S.~Han, ``Haq: Hardware-aware automated
  quantization with mixed precision,'' in \emph{Proceedings of the IEEE/CVF
  Conference on Computer Vision and Pattern Recognition}, 2019, pp. 8612--8620.

\bibitem{zhang2018lq}
D.~Zhang, J.~Yang, D.~Ye, and G.~Hua, ``Lq-nets: Learned quantization for
  highly accurate and compact deep neural networks,'' in \emph{Proceedings of
  the European conference on computer vision (ECCV)}, 2018, pp. 365--382.

\bibitem{DLMC}
\BIBentryALTinterwordspacing
{Google Research. [n.d]}, ``Deep learning matrix collection,'' 2021. [Online].
  Available:
  \url{https://github.com/google-research/google-research/tree/master/sgk}
\BIBentrySTDinterwordspacing

\bibitem{dosovitskiy2020image}
A.~Dosovitskiy, L.~Beyer, A.~Kolesnikov, D.~Weissenborn, X.~Zhai,
  T.~Unterthiner, M.~Dehghani, M.~Minderer, G.~Heigold, S.~Gelly \emph{et~al.},
  ``An image is worth 16x16 words: Transformers for image recognition at
  scale,'' \emph{arXiv preprint arXiv:2010.11929}, 2020.

\bibitem{yuan2021tokens}
L.~Yuan, Y.~Chen, T.~Wang, W.~Yu, Y.~Shi, Z.-H. Jiang, F.~E. Tay, J.~Feng, and
  S.~Yan, ``Tokens-to-token vit: Training vision transformers from scratch on
  imagenet,'' in \emph{Proceedings of the IEEE/CVF International Conference on
  Computer Vision}, 2021, pp. 558--567.

\bibitem{liu2021swin}
Z.~Liu, Y.~Lin, Y.~Cao, H.~Hu, Y.~Wei, Z.~Zhang, S.~Lin, and B.~Guo, ``Swin
  transformer: Hierarchical vision transformer using shifted windows,'' in
  \emph{Proceedings of the IEEE/CVF International Conference on Computer
  Vision}, 2021, pp. 10\,012--10\,022.

\bibitem{arnab2021vivit}
A.~Arnab, M.~Dehghani, G.~Heigold, C.~Sun, M.~Lu{\v{c}}i{\'c}, and C.~Schmid,
  ``Vivit: A video vision transformer,'' in \emph{Proceedings of the IEEE/CVF
  International Conference on Computer Vision}, 2021, pp. 6836--6846.

\bibitem{tay2020long}
Y.~Tay, M.~Dehghani, S.~Abnar, Y.~Shen, D.~Bahri, P.~Pham, J.~Rao, L.~Yang,
  S.~Ruder, and D.~Metzler, ``Long range arena: A benchmark for efficient
  transformers,'' \emph{arXiv preprint arXiv:2011.04006}, 2020.

\bibitem{MI200}
AMD, ``{AMD Instinct MI200 Instruction Set Architecture Reference Guide},''
  2022.

\bibitem{megatron}
\BIBentryALTinterwordspacing
D.~Narayanan, M.~Shoeybi, J.~Casper, P.~LeGresley, M.~Patwary, V.~Korthikanti,
  D.~Vainbrand, P.~Kashinkunti, J.~Bernauer, B.~Catanzaro, A.~Phanishayee, and
  M.~Zaharia, ``Efficient large-scale language model training on {GPU}
  clusters,'' \emph{SC}, 2021. [Online]. Available:
  \url{https://arxiv.org/abs/2104.04473}
\BIBentrySTDinterwordspacing

\bibitem{li2021chimera}
S.~Li and T.~Hoefler, ``Chimera: efficiently training large-scale neural
  networks with bidirectional pipelines,'' in \emph{Proceedings of the
  International Conference for High Performance Computing, Networking, Storage
  and Analysis}, 2021, pp. 1--14.

\bibitem{shoeybi2019megatron}
M.~Shoeybi, M.~Patwary, R.~Puri, P.~LeGresley, J.~Casper, and B.~Catanzaro,
  ``Megatron-lm: Training multi-billion parameter language models using model
  parallelism,'' \emph{arXiv preprint arXiv:1909.08053}, 2019.

\bibitem{zheng2022alpa}
L.~Zheng, Z.~Li, H.~Zhang, Y.~Zhuang, Z.~Chen, Y.~Huang, Y.~Wang, Y.~Xu,
  D.~Zhuo, J.~E. Gonzalez \emph{et~al.}, ``Alpa: Automating inter-and
  intra-operator parallelism for distributed deep learning,'' \emph{arXiv
  preprint arXiv:2201.12023}, 2022.

\bibitem{pauloski2021kaisa}
J.~G. Pauloski, Q.~Huang, L.~Huang, S.~Venkataraman, K.~Chard, I.~Foster, and
  Z.~Zhang, ``Kaisa: an adaptive second-order optimizer framework for deep
  neural networks,'' in \emph{Proceedings of the International Conference for
  High Performance Computing, Networking, Storage and Analysis}, 2021, pp.
  1--14.

\bibitem{osawa2019large}
K.~Osawa, Y.~Tsuji, Y.~Ueno, A.~Naruse, R.~Yokota, and S.~Matsuoka,
  ``Large-scale distributed second-order optimization using kronecker-factored
  approximate curvature for deep convolutional neural networks,'' in
  \emph{Proceedings of the IEEE/CVF Conference on Computer Vision and Pattern
  Recognition}, 2019, pp. 12\,359--12\,367.

\end{thebibliography}

\end{document}